\documentclass[showpacs,showkeys,prb,twocolumn]{revtex4}
\usepackage{graphicx,graphics}
\usepackage{supertabular}
\usepackage{longtable}
\usepackage{bm}
\usepackage{amsmath}

\begin{document}

\title{Paraxial Hamiltonian for photons in two-dimensional photonic crystal microstructures}
\author{D. L. Boiko}
\email{dmitri.boiko@epfl.ch} \affiliation{Quantum Architecture
group, \'Ecole Polytechnique F\'ed\'erale de Lausanne, 1015,
Lausanne, Switzerland
\flushleft{\textnormal{(Received:~}}\\
}

\begin{abstract}
New solid-state physics based approach is developed for analysis
of the paraxial light propagation in two-dimensional (2D) photonic
lattices of coupled dielectric waveguides or microcavities. In
particular, using Maxwell's equations, a non-Hermitian Hamiltonian
eigenproblem with respect to the spinor wave function of a photon
is obtained for energy-dissipating photonic microstructures. The
Hamiltonian is suitable for almost the entire subclass of 2D
structures encompassing arrays of semiconductor microcavities and
microstructured photonic crystal fibers, characterized by light
propagating mostly normal to the periodic lattice plane. Methods
of numerical solution are discussed and the formalism is applied
to a square array of coupled semiconductor microcavities,
revealing reach possibilities for tailoring photonic band
structure both in the photon energy and photon lifetime energy
broadening domains. In particular, a feasibility to open a double
photonic crystal band gap simultaneously in the energy and
lifetime energy broadening domains is demonstrated.
\end{abstract}

\date{\today}
 \maketitle

\section{ Introduction}
Photonic crystal structures offer unique possibilities for
controlling light-matter interactions\cite{Kogelnik71, Bykov72,
Yablonovitch87,Akahane03} and tailoring 
light propagation\cite{Mekis96}
by introducing 
lattice defects\cite{John87,Foresi97} or photonic crystals
heterostructure barriers.\cite{Yano01,Guerrero04} 
Among 
proposed so far photonic structure configurations, two-dimensional
(2D) 
lattices 
are highly attractive in 
technological
aspect 
and have a realistic potential of finding applications in novel
optoelectronic devices and integrated photonic circuits.

A particular subclass of 2D structures consists of arrays of
coupled optical waveguides or microcavities. In these structures,
a
photonic mode propagates mostly along the 
waveguiding direction or cavity axis [vertical direction in
Fig.\ref{fig1} (a)], such that only a small lateral 
component $\mathbf{k}_\perp$ of wave vector $\mathbf{k}$ undergoes
Bragg reflections. Such paraxial
photonic structures employ 
lattices of period significantly exceeding the optical
wavelength.
Photonic crystal fibers\cite{Russell03} are an example of
implementing this concept. Matrices of phase-coupled vertical
cavity surface emitting lasers (VCSELs)
are another example of such 2D structures.\cite{Orenstein91,
Morgan92} These structures allow photonic wave functions
behaviour to be directly examined in periodic and quasi-periodic
lattices\cite{Pier97,Pier00} and offer a possibility to implement
photonic crystal heterostructures made of 
several photonic crystal materials with dissimilar 
band gaps.\cite{Guerrero04, Lundeberg05} 
In addition,
they might 
be 
loaded with optical gain or loss, which are shown\cite{McGurn93,Kuzmiak94,Sigalas95} 
to impact 
band edge energies 
at 
high-symmetry points of reciprocal lattice.

Many of these concepts related to photonic crystal structures
originate from analogy to semiconductors (see for example,
Refs.~\onlinecite{Boiko06C} and \onlinecite{Boiko07}
in 
the case of author).
However, despite an apparent simplicity of peculiar 
2D photonic crystal microstructures presented 
above,
a simple 
approach utilizing 
a standard Hamiltonian framework
of the solid-state physics 
has not yet been reported so far for photonic crystals.

In this paper, 
a single-photon non-Hermitian Hamiltonian 
is obtained for 
2D photonic lattices 
of parallel 
dielectric waveguides or microcavities. It is derived 
by
introducing a 
paraxial gauge transformation, which
converts
Maxwell's equations 
into
a non-Hermitian
Hamiltonian eigenproblem with respect to a 
biorthonormal set of 
spinor 
wave functions for photons. (See Eq.~(\ref{H_all}), the main
result of this paper.) 
Properties of this Hamiltonian
in case of lattices 
with inversion symmetry (\textit{e.g.},  widely used square and
triangular lattices) are discussed and a method of numerical
solution based on biorthonormal plane wave expansion is outlined.
The numerical
solutions 
are obtained for the structures with square symmetry of the
lattice, predicting 
a peculiar interplay of the photon energy and photon lifetime
energy broadening bands. These results 
envisage 
new 
possibilities for application of dissipative 
photonic crystals
heterostructures 
benefiting from 
the features of $2N$-dimensional confinement of photonic envelop
wave functions in
$N$-dimensional photonic lattices. 



For the first time, a truncated form of this Hamiltonian (adapted
for the case of VCSEL arrays)
has been
briefly 
introduced
in Ref.~\onlinecite{Boiko02}, without %
providing a 
discussion about 
its validity 
or proving
orthogonality 
of its
solutions. However, 
the symmetry and polarization structure of the main lasing
modes predicted by the model 
has been confirmed in Ref.~\onlinecite{Boiko02} by experimental
measurements in square arrays of VCSELs.
(See also Refs.~\onlinecite{Boiko04,Guerrero04B}.)
%
%
A $\mathbf{k{\cdot}p}$ approximation of this Hamiltonian restricted to the eight 
bands\cite{Boiko07} 
has been 
solved 
%
%
analytically, 
yielding the dispersion characteristics of the low-order bands in
square lattices.\cite{Boiko08}
%
%
A 
parabolic 
approximation 
obtained for the optical loss dispersion 
allows
a single-band effective mass Hamiltonian to be introduced for
modeling properties of the main lasing modes in VCSEL arrays.
%
%
%
Its finite difference element
representation  
with the numerical grid 
step being the array 
pitch
has the form of 
Coupled Mode Theory equations, offering an 
efficient tool 
for calculating 
photonic envelope functions in complex
quasiperiodic 
lattices.\cite{Guerrero04}
%
This method, 
which is 
detailed in the later Ref.~\onlinecite{Lundeberg07},
%
%
has been 
%
%
used to interpret the experimental measurements of confined
optical envelop functions in VCSEL-based photonic crystal
heterostructures. \cite{Guerrero04,Lundeberg05} 

%
%

%
%
In Ref.\onlinecite{Boiko07}, this Hamiltonian
is used
to analyze the Coriolis-Zeeman effect for photons in periodic
lattices of microcavities, envisaging possibility of interactions
between photons and gravitational field.
The basic steps of analysis 
in Ref.~\onlinecite{Boiko07}
might be used as guidelines for following
the treatment of photonic lattices 
in the present paper. However, note that in
Ref.~\onlinecite{Boiko07},
the dissipative effects 
are not taken into account.

The paper is organized as follows.
In Sec.~\ref{SecMirrorPatterned}, using an equivalent
cavity-unfolded representation of microcavities,
the Hamiltonian for lattices of microcavities defined by mirror
reflectivity patterning 
is obtained.
In Sec.~\ref{SecIndexPatterned},
the 
Hamiltonian is extended to the case of 
microcavities or parallel dielectric waveguides defined by
periodic
variations of refractive index. 
Sec.~\ref{ResultsAndDiscussion} brings together 
the non-Hermitian Hamiltonian terms
obtained in previous sections, discuss 
the 
validity of such 
Hamiltonian approximation, outlines the method of solution based
on biorthonormal plane wave expansion 
and reports numerical results of band structure computations in
square-lattice 
photonic crystal microstructures.

%
%

\section{Reflectivity-patterned microcavities}
\label{SecMirrorPatterned}

The Hamiltonian for 
lattices of microcavities with 
periodic variations 
of 
mirror reflectivity is detailed 
here on example of
mirror-patterned VCSEL
arrays (Fig.~\ref{fig1}).
%
In Sec.~\ref{VCSELstructure}, a typical structure of
 VCSEL array with
reflectivity-patterned distributed Bragg reflector
(DBR) is introduced. 
In Sec.~\ref{SubSecMatEq},
the constitutive equations in cavity-unfolded representation are
obtained. 
Sec.~\ref{SubSecHamiltonian} reports
the 
model Hamiltonian. 

\subsection{VCSEL array photonic crystal}


\begin{figure}[tbp]
\includegraphics{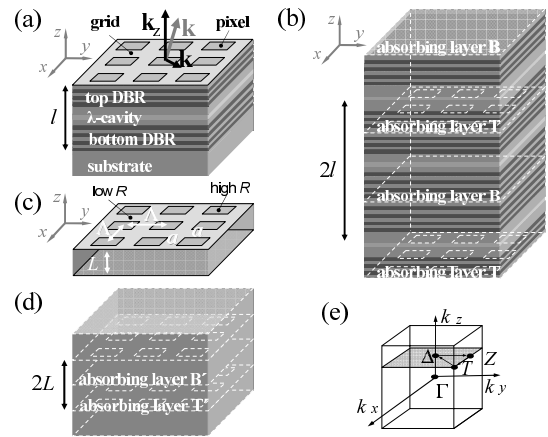}
\caption{VCSEL array photonic crystal 
(a) Schematic illustration of the
wafer structure and metal-patterned top DBR composition. (b)
Cavity-unfolded representation of VCSEL array 
with equivalent (absorbing) $\mathbf{g}$-layers 
representing the effect of reflections at 
the
top (T) and bottom (B) 
cavity interfaces. (c) Simplified model system consisting of a
Fabry-P\'erot cavity with reflectivity-modulated mirror. (d)
Cavity-unfolded representation of the model system in (c). (e)
Brillouin zone of the cavity-unfolded photonic crystal.
$\Lambda$ is the lattice pitch, $a$ is
the width of square VCSEL pixel.} \label{fig1}
\end{figure}

\label{VCSELstructure}  Arrays of vertical cavity surface
emitting lasers are an example of photonic crystal made of
evanescently coupled microcavities. These particular
photonic 
structures utilize a usual VCSEL wafer
incorporating 
a one-wavelength 
cavity 
with an optical gain medium (e.g. quantum wells) sandwiched
between two DBRs 
[Fig.\ref{fig1}
(a)].
The $\lambda $-cavity
might be regarded as a defect in the
thick $\lambda /2$-periodic DBR stack.
In a solitary VCSEL, it defines the lasing mode wavelength, which
fits the cavity roundtrip self-repetition conditions. In VCSEL
arrays, the $\lambda $-cavity fixes the main (longitudinal) wave
vector component [along the $z$-axis in Fig.\ref{fig1}(a)].

The photonic lattice of VCSEL array is defined by reflectivity
patterning, $R(x,y)$,  of
the top DBR.
\cite{Orenstein91} The crystal unit cell consists of a highly
reflecting VCSEL pixel surrounded by a grid of lower reflectivity.
In the bottom-emitting structures, this is accomplished by
depositing a periodic pattern of metallic overlays on the top DBR, \emph{e.g.}, 
using gold pixels and chromium grid for pattern
definition.\cite{Orenstein91}
In the top-emitting structures, the pixel positions and the grid
are defined by air openings in the metallic film (\emph{e.g.},
 gold film) deposited on the
top DBR
.\cite{MontidiSopra00} The pattern is characterized in terms of a
reflectivity contrast and 
lattice cell fill factor $FF$ (
the
area ratio of 
the 
pixel and
of the lattice 
cell). The resulting composite DBR is of high reflectivity at the
pixel positions ($R ^2 {\sim} 99.9\% \div 99.99 \%$) and of low
reflectivity contrast between the pixels and the grid ($2 \delta
R {\sim}1\%$). Since the roundtrip optical gain is low and
uniformly distributed across the VCSEL structure,
%
%
such shallow reflectivity pattern
suffices to define position of the lasing microcavities.

For a typical $InGaAs/GaAs/AlGaAs$ VCSEL array emitting at $960$
$nm$ wavelength, the lattice pitch $\Lambda$ is about 5 $\mu m$,
such that only a small transversal component $k_{\bot }{\sim} \pi
/ \Lambda $ of the propagation vector $\mathbf{k}$ ($k_{\bot
}/k\sim 0.03$)  undergoes Bragg reflections in the plane of
periodic lattice. The angular spectrum of the most important
low-order cavity modes is thus of narrow width $\xi\sim k_{\bot
}/k$ and centered about the main propagation direction (the
$z$-axis direction).
In what follows, the cavity modes 
are analyzed in the paraxial approximation, within an accuracy of
second-order terms 
$\xi^2{\sim} 10^{-3}$.
The impact of reflectivity pattern $R(x,y)$ is taken into account
as a second-order perturbation. (For a typical VCSEL array, the
amplitude reflectivity variations 
assume $\delta R \alt 0.5 \cdot 10^{-3}$.)

The optical gain is uniformly distributed across the
cavity\cite{Boiko06B} and has no influence on the cavity mode
structure. 
Uniformly distributed optical losses, such as the material losses
($\sim 0.1\%$ in VCSEL arrays) and the output coupling losses at
the bottom DBR 
($\sim 0.16\%$), do not
influence 
a 
curvature of the dispersion curves 
for the cavity modes as well. 
Their additive contribution to the optical mode frequencies and
losses is ignored in the subsequent analysis and 
can be easily taken into account 
by shifting the frequency and loss spectra %
of
optical modes.

An optical mode of VCSEL array can be regarded as a superposition
of standing waves 
coupled via Bragg scattering effects
at the patterned mirror.
The partial amplitudes of 
standing-wave harmonics are defined thus by the boundary
conditions at
the cavity mirrors 
and not as a 
result of propagation within the structure, like in usual
photonic crystal materials exhibiting 
periodic variations of
the dielectric constant.
%
%
%
%
%
%
To apply the usual method of orthogonal plane wave (OPW)
expansion,\cite{Leung90}
an equivalent cavity-unfolded representation is needed. 

Multiple reflections at the cavity mirrors effectively
translate the entire cavity into a 
structure that is periodic along the cavity axis ($z$-axis). The
unfolded PhC is three-dimensional (3D) and 
can be analyzed in
terms of propagating plane waves. Optical modes of a VCSEL array
are thus represented by electromagnetic Bloch waves propagating in
the equivalent 3D photonic crystal.

A typical VCSEL wafer incorporates a few tens of various 
dielectric material layers. 
To take into account a
detailed 
composition of the wafer structure, the cavity has to be unfolded
using the outermost layer interfaces. Thus reflections at the
metallic overlay pattern in the top DBR and at the wafer
substrate in the bottom Bragg reflector have to be used for
unfolding the VCSEL cavity [Fig.\ref{fig1}(a)].


%


%
%

At each reflection from cavity 
output coupling interfaces,
the field amplitude 
reduces 
because of the waves leaving the cavity. In unfolded 
photonic crystal, 
the impact of subsequent reflections is reproduced by a periodic
stack of energy dissipating layers ($\mathbf{g}$-layers), each
representing the effect of a single mirror reflection
[Fig.\ref{fig1}(b)]. %
%
The zero propagation length
between the incident and reflected waves at the output coupling
interfaces of the cavity assumes that the equivalent 
$\mathbf{g}$-layers are infinitesimally thin.
Introduction of such 
thin absorbing layers allows both the energy dissipation (within
the layers) and the continuity of electromagnetic field components
(at each layer interface) to be taken by the model into account.

In between the 
$\mathbf{g}$-layers, the unfolded
photonic crystal has the same sequence of dielectric layers
as encountered in the VCSEL microcavity.
%
%
%
%
%
%
%
%
However, such a
detailed description 
of 
wafer structure
composition is needed for 
analyzing 
technical issues, 
like, \emph{e.g.}, optimizing a spectral 
overlap between the
gain peak and the cavity modes in function of
the 
active region temperature.\cite{Boiko06B}
This description 
leads to 
intricate numerical model demanding lengthy numerical simulations.

This paper is focused on general features 
of the light propagation behaviour in photonic lattices.
Therefore,  a simplified model system is
used\cite{Boiko02,Boiko04}
[Fig.\ref{fig1}(c)]. It consists of the $\lambda$-cavity of VCSEL array 
in which the DBRs 
are replaced with ideal mirrors of effective DBR 
reflectivities. In this way, a Fabry-P\'erot cavity
is formed in which
the reflectivity of
the upper 
mirror is modulated in two directions parallel
to its plane.
Such Fabry-P\'erot resonator
replicates
the boundary conditions for the field in VCSEL cavity,
reproducing thus the effect of fast
oscillations
within the periodic DBR stacks.
Introduction of this model system is consistent with
the Kapitza method\cite{LandauI} to analyze 
oscillations in a dynamic system. In this particular case, it
consists in separating fast longitudinal oscillations of the
electromagnetic field (along the cavity axis) and its slow lateral
oscillations
(in the $x$-$y$ plane).
%
%

The model system treated here consists thus of a Fabry-P\'erot
cavity of
the length $L{=}\lambda / n$ with patterned reflectivity $R(x,y)$
of one mirror [the upper mirror in Fig.\ref{fig1}.(c)]. The
second cavity mirror is of uniform reflectivity and does not
impact the cavity mode structure. In what follows, 
it is assumed to be perfectly reflecting. The dielectric material
in the cavity is of uniform refractive index $n$ and
impedance $Z$. 
The
equivalent cavity-unfolded representation 
is a 
simple
$2L$-periodic (along the $z$ axis) structure shown 
in Fig.\ref{fig1}(d).

%
%
%
%
%
Note that 
the periods (along the $z$ axis) of the
cavity-unfolded VCSEL array structure [Fig.\ref{fig1}(b)] and its
simplified model
[Fig.\ref{fig1}(d)] are different.
%
%
%
As shown below, the 
absorbing $\mathbf{g}$-layers 
have 
the same impedance as
the neighboring dielectric materials 
such that 
Bragg reflections in the $z$-axis
direction cannot occur at the $g$-layers. 
Therefore, 
the 
period 
of the cavity-unfolded structure has no impact on
photonic bands. 

Furthermore, the longitudinal wave vector component of a mode is set 
by the self-repetition condition at the cavity roundtrip.
However, in the cavity-unfolded representation, it
is not defined 
until such additional condition 
is superimposed, yielding $k_z=2\pi/L$.

\subsection{Material equations for a cavity-unfolded structure}
\label{SecMatEq} \label{SubSecMatEq}

There are many different
conventions for the coordinate systems used to consider
reflections, and the phases of reflection coefficients are
dependent on the coordinate system. Here, the following
convention is used for the complex vector amplitudes of the
incident and reflected waves:
\begin{equation}
\left[
\begin{array}{l}
\mathbf{E} \\
\mathbf{H}%
\end{array}%
\right]_{s,p}^{(r)}=R_{s,p}\hat{\sigma} \left[
\begin{array}{l}
\mathbf{E} \\
\mathbf{H}%
\end{array}%
\right] _{s,p}^{(i)} \label{reflection}
\end{equation}%
where $\hat{\sigma}$ is the reflection operator, $R_{s,p}$ are the
amplitude reflection coefficients, indexes $s$ and $p$ stand for
polarization perpendicular to the plane of incidence and in the
plane of incidence, respectively. The convention
(\ref{reflection}) yields equal (in modulus and phase) reflection
coefficients $R_{s}$ and $ R_{p}$ at normal incidence,
$R_{s}=R_{p} (=R )$. Furthermore, in the range of propagation
angles $\xi{=}k_\perp/k_z$ considered here, 
%
%
the angular dispersion of the DBR reflectivity is negligible,
such that the approximation $R_{s}{=}R_{p}$ is valid at oblique
incidence as well.
Indeed, for 
$\xi {=} 0.03$, the reflectivity
of a typical DBR
decreases on $\delta R {\sim} 10^{-5}$. 
(The reflectivity amplitude at normal incidence is $R
{\sim} 1$.) 


The notion of
absorbing $\mathbf{g}$-layers in the cavity-unfolded
representation of a structure is linked to the duality
of symmetry
operations.\cite{Altmann65} In the active notation, 
the reflection operator $\hat{\sigma}$ in (\ref{reflection})
assumes transformation of the vectors $\mathbf{E} (x,y,z)$ and
$\mathbf{H}(x,y,z)$.
In the passive notation this 
symmetry operation is applied to the coordinate axes. 
The effect of $R\hat{\sigma}$ on the field 
can be then regarded 
as a result of propagation
through a
layer with 
equivalent transmittance 
$T{=}R$. 
To match the phases 
of reflected (at the cavity mirror) and transmitted (by the
equivalent $g$-layer) waves, the layer thickness has to be
infinitesimally small. 
To impact the amplitude and
phase of the transmitted waves in this limit, 
the material parameters of the layer should vary as $\propto
h^{-1}$, with $h$ being the layer thickness.


In the cavity-unfolded representation, 
electromagnetic waves propagate in the periodic stack of
absorbing $\mathbf{g}$-layers 
[Fig.\ref{fig1}(d)]. Material equations for this
structure are obtained here by considering the 
amplitude
of a plane
electromagnetic wave in 
Fabry-P\'erot (FP) cavity with 
mirrors of uniform reflectivity.
%
%
%
The cavity mirror located at $z{=}0$
is of 
reflectivity $|R|{<}1$. As in Sec.~\ref{VCSELstructure},
the second cavity mirror (at $z{=}{-}L$) is a perfect reflector.
In the cavity-unfolded representation, 
the
amplitude of
the 
wave traveling towards the positive $z$-axis direction reduces by
a factor of $R$ at each 
$g$-layer:
\begin{equation}
\left[
\begin{array}{l}
\mathbf{E} \\
\mathbf{H}%
\end{array}%
\right] =e^{i\mathbf{kr}-i\omega t}  R^{ \sum_{j}\left[ \theta
(z-2jL)-\frac{1}{2}\right] -\frac{z}{2L}} \left[
\begin{array}{l}
\mathbf{E}_{0} \\
\mathbf{H}_{0}%
\end{array}%
\right] ,\label{EinFP_off_ax}
\end{equation}%
where $\omega$ is the frequency of electromagnetic wave,
$\mathbf{E}_{0}$ and $\mathbf{H}_{0}$ are the amplitudes of
electric and magnetic fields at the coordinate origin $z{=}0$, $\theta (z)$ is the 
unit step function
%
%
$\theta (z){=}\left\{ 0~
(z{<}0), ~1/2 ~
(z{=}0),~1~
(z{>}0) \right\} $, and  $j$ is an integer number
($j{=}0,\pm1,\ldots$)
enumerating 
$g$-layers 
(at $z{=}0,\pm2L, \ldots$) 
associated with the
subsequent reflections at the 
cavity mirror.

The electromagnetic wave 
(\ref{EinFP_off_ax})
is written 
in the form of the Bloch wave composed of plane wave envelope
function and periodic crystal part (the second exponential term is
of the period $2L$).
%
The effective propagation vector $\mathbf{k}$ of the wave
(\ref{EinFP_off_ax}) is complex:
\begin{equation}
\mathbf{k}=\bm{\tau}\frac{\omega n
}{c}-i\mathbf{\hat{z}}\frac{\ln R}{2L}, \label{Kpropag}
\end{equation}%
where $\omega n/c$ and $\bm\tau$ are the wave number and unit
vector along the wave
propagation direction 
as seen
from the cavity, and $\mathbf{\hat{z}}$ is the $z$-axis unit
vector.
%
%
%
%
%
%
%
Eqs.(\ref{EinFP_off_ax})-(\ref{Kpropag}) 
show 
the evolution
of 
wave amplitude and phase with 
successive roundtrips in the cavity.
The imaginary part of $\mathbf{k}$ accounts for the decay of
electromagnetic field in the cavity due to 
output coupling 
loss.
The real part of $\mathbf{k}$ allows 
the cavity mode 
frequencies to be evaluated 
from a usual condition ${\text{Re}}\left(2 k_{z} L \right){=} 2
\pi q$ with $q$ being an integer.

In the cavity [Fig.\ref{fig1}(c)], and hence
in between the $\mathbf{g}$-layers of unfolded photonic crystal
[Fig.\ref{fig1}(d)], the electric and magnetic field components
(\ref{EinFP_off_ax}) assume the relationships
\begin{equation}
\bm{\tau}\mathbf{\times E} =Z \mathbf{%
H,\qquad } \bm{\tau}\mathbf{\times H} =-Z^{-1}\mathbf{E}.
\label{E0&H0in_TE}
\end{equation}%
%
%
%
%
%
%
%
A correspondence between the electromagnetic energy
flux in the 
cavity and in the equivalent unfolded structure 
can be established assuming that
%
%
the impedance 
of $\mathbf{g}$-layer material is the
same as in 
the dielectric material of the cavity. 
Otherwise, an impedance mismatch and Bragg scattering effects at periodic stack of $\mathbf{g}$-layers 
will 
result in %
a retro-reflected
wave propagating in the backward $z$-axis direction, having no
counterpart in the FP cavity.
%
%
The relationships (\ref{E0&H0in_TE}) thus hold
through the entire equivalent photonic crystal, including the
absorbing $\mathbf{g}$-layers as well.

%
%



%
%
Substituting (\ref{EinFP_off_ax})-(\ref{E0&H0in_TE}) in Maxwell'
equations
%
%
\begin{equation}
\begin{split}
div\mathbf{B} &= 0,\qquad rot\mathbf{E}=-\frac{1}{c}\frac{\partial \mathbf{B}%
}{\partial t},   \\
div\mathbf{D} &= 0,\qquad rot\mathbf{H}=\frac{1}{c}\frac{\partial \mathbf{D}%
}{\partial t},  
\end{split} \label{MaxwellEQS}
\end{equation}%
%
within the accuracy of the $\xi^2$-order terms, one obtains the
constitutive relationships for the 
periodic 
structure representing 
cavity-unfolded Fabry-Perot
resonator 
%
%
\begin{eqnarray}
\mathbf{D}&=&\varepsilon \mathbf{E}+ \mathbf{H\times g} , \quad
\mathbf{B}=\mu \mathbf{H}+ \mathbf{g\times E}, \label{MatEqED} \\
\mathbf{g}&=&- i\mathbf{\hat{z}} \frac{c \ln R}{\omega }
\sum_{j}\delta (z-2jL), \label{MatEqs}
\end{eqnarray}%
%
%
where $\varepsilon {=} n/Z $ and $\mu {=} n Z$ are the dielectric
constant and magnetic permeability in the cavity,
$\int_{-\infty}^z \delta (\xi)d\xi {=} \theta(z)$.
Eqs.~(\ref{MatEqED})-(\ref{MatEqs}) assume the impedance matching
through the entire 
structure, which
%
thus exhibits 
no artificial photonic band gaps
in the $z$-axis direction.

%
%

Eqs. (\ref{MatEqED})-(\ref{MatEqs}) are similar to the
constitutive equations in a
nonpermanent gravitational field 
induced by 
rotating coordinate frame.\cite{Heer64,Khromykh66,Post67} Indeed, in Eq.~(\ref{reflection}), 
the reflection operator $\hat{\sigma}$ at the cavity mirror
includes the coordinate rotation (
$\hat{\sigma}=\hat{I} \hat{C}_{2}(\mathbf{\hat{z}})$ where
$\hat{I}$ represents the coordinate inversion and
$\hat{C}_{2}(\mathbf{\hat{z}})$ represents rotation by $\pi $
about the cavity axis). The non-Galilean space-time
metric\cite{LandauII} ($\mathbf{g}\neq0$) within the 
layers represents 
the effect of mirror reflections and can thus be attributed to the
coordinate 
rotations used to unfold the cavity.

%
An apparent discontinuity of the electromagnetic field at
$\mathbf{g}$-layers [see Eqs.(\ref{EinFP_off_ax}) and
(\ref{MatEqs})] is in fact
%
%
a result of infinitesimally small $\mathbf{g}$-layer thickness.
%
%
Using Eq.~(\ref{MatEqs}) and approximation $\delta (z){=}\lim_{ h
\rightarrow 0}\frac{1}{h}\left\{ \theta \left( z{+}\frac{
h}{2}\right) {-}\theta \left( z{-}\frac{h}{2}\right) \right\} $,
one can obtain that
\begin{equation}
\mathbf{g}=-i\mathbf{\hat{z}}\frac{c\ln R}{\omega }h^{-1}
\label{MatEq_eqvLayerH}
\end{equation}%
%
%
%
in a layer of small 
finite
thickness $h$ (\emph{e.g.}, located at the coordinate origin). 
It can be seen 
that 
the tangential fields
$\mathbf{E}$ and $\mathbf{H}$ as well as the normal components of
$\mathbf{D}$ and
 $\mathbf{B}$ are 
 continuous at each interface of the $\mathbf{g}$-layer,
in agreement with the
boundary conditions at moving (rotating) interfaces.
\cite{Bolotovskii74,
Bolotovskii89, LandauVIII} 
Within the 
layer, the tangential components of $\mathbf{D}$ and $\mathbf{B}$
vary as $\propto h^{-1}$, yielding 
a singularity in the limit 
$h {\rightarrow} 0$.

The effective refractive index in 
the 
layer 
varies
with the
wave propagation direction\cite{Khromykh66,Boiko98,Boiko07} $\bm{\tau}$: 
\begin{equation}
n_{\text{eff}}=n+\bm{\tau}\mathbf{g}. \label{index_layer}
\end{equation}
For a plane wave propagating 
in the positive
$z$-axis direction ($\mathbf{\hat{z}}{\cdot}\bm{\tau}{>}0$), 
the phase accrual
in the layer
%
%
reads $\int \frac{\omega }{c}n_\text{eff}  \bm{\tau}
d\mathbf{r}$, yielding a
complex number
\begin{eqnarray}
\phi =
\int_{-h /2}^{h/2}\frac{\omega }{c}\left(
\mathbf{\hat{z}g}\right) dz =\mathbf{-}i\ln R, \label{Phase}
\end{eqnarray}%
where all other terms vanish in the limit $h {\rightarrow} 0$.
At 
each
$\mathbf{g}$-layer, the wave amplitude thus reduces by a factor
of $e^{i\phi}=R$,
%
%
in agreement with Eq.~(\ref{EinFP_off_ax}).
%

%
The wave propagation
 in the positive
$z$-axis direction of the 
structure 
(\ref{MatEqED})-(\ref{MatEqs})
corresponds to 
multiple roundtrips in the FP cavity.
%
%
However, this 
periodic structure 
is nonreciprocal.
Thus, the amplitude of a wave propagating 
in the
backward direction ($\bm{\tau}\mathbf{\hat{z}}<0$) increases by a
factor of $R^{-1}$  (where $|R|{<}1$) at 
each $\mathbf{g}$-layer. 
This picture 
corresponds to 
an external electromagnetic field
exciting oscillations in the FP cavity.
The opposite $z$-axis
directions in the cavity-unfolded structure
(\ref{MatEqED})-(\ref{MatEqs})
are thus 
related by 
the 
time reversal operation.

In the case of mirrors with 
uniform reflectivity, 
the vector $\mathbf{g}$
in 
(\ref{MatEqs}) shows the same transformation properties 
as 
vector $\mathbf{g}=\frac 1c \mathbf{\Omega} {\times} \mathbf{r}$
composed of space-time components of metric tensor in the case of
rotations. Namely, $\mathbf{g}$ is anti-invariant ($\mathbf{g}
\rightarrow -\mathbf{g}$) under the coordinate inversion ($P$)
 and time reversal ($T$) operations.
It can be seen that Eqs.(\ref{MatEqED})-(\ref{MatEqs}) do not
show 
any particular symmetry under the $PT$-transformation (coordinate
inversion followed by time reversal operation). Therefore,
 the  energy spectrum assumes complex eigenvalues, as opposed to
the real spectrum in the case of pseudo-Hermitian systems  that
are invariant under the $PT$-transform.\cite{Bender98,Mostafazadeh02}
%
However,  the $PT$
transformation is important 
for the analysis presented here. Thus, in 
Sec.
\ref{ResultsAndDiscussion}, square-lattice 
structures are treated using 
non-Hermitian Hamiltonian and biorthonormal set of wave function
partners.
The invariance of square lattice under (2D) coordinate inversion
allows the partners
of 
biorthonormal set to be defined as 
$\Psi(t,x)$ and $\tilde{\Psi}(t,x)=\Psi^{*}(-t,-x)$.
(In the case of
dissipative effects,
the results
of 
substitution 
$t \rightarrow -t$ and 
complex conjugation
are different,\cite{Kogelnik75} such that a wave function
$\Psi(t,x)$ and its $T$-transform 
$\Psi^{*}(-t,x)$ do not coincide.)

\subsection{Hamiltonian}
\label{SubSecHamiltonian}

%

The material equations (\ref{MatEqED})-(\ref{MatEqs}) define
a
cavity-unfolded 
periodic structure, which is equivalent to a Fabry-P\'erot
resonator with mirrors of uniform reflectivity. 
The 
vector $\mathbf g$ in (\ref{MatEqs}) does not vary with the
position in the $x$-$y$ plane (parallel to the cavity mirrors).
Different from FP cavity, 
lattices of coupled microcavities, such as 
VCSEL arrays
discussed here, utilize mirrors with 
reflectivity patterning $R(x,y)$. The cavity-unfolded
representation of such mirror-patterned structures is assumed
to obey the same relationships
(\ref{MatEqED}), however, with periodically varying (in the
$x$-$y$ plane) space-time
coupling 
%
%
%
\begin{equation}
\mathbf{g}=- i\mathbf{\hat{z}}\frac{c \ln R(x,y)}{\omega
}\sum_{j}\delta (z-2jL), \label{3Dg}
\end{equation}%
where
the z-period of unfolded structure 
is $2L$,
and $R(x,y)$ is the 
(amplitude) reflectivity 
of the cavity mirror
(\emph{e.g.}, effective reflectivity of the top DBR in VCSEL
arrays).
%
%
%
Note that in the model
defined by Eqs.(\ref{MatEqED}) and (\ref{3Dg}), like in the case of 
other photonic
crystal structure models,
the 
field radiation effects at the
boundary discontinuities\cite{Kurilko68} are neglected.




It can be seen that 
an
electromagnetic Bloch wave propagating in 
photonic crystal with 
effective 
noninertiality (\ref{3Dg}) is of the
same form 
as indicated 
in Eq.~(\ref{EinFP_off_ax}) but 
the field amplitudes $\mathbf{E}_{0}$ and $\mathbf{H}_{0}$ vary
periodically in the $x$-$y$ plane.
%
%
%
%
%
%
%
%
%
%
%
In 
(\ref{EinFP_off_ax}), the frequency $\omega $
(real number) is 
the
independent parameter of motion, while 
the Bloch vector $\mathbf{k}$ (complex)  
is defined 
from the dispersion equation 
(\ref{Kpropag})
and takes
the effective propagation loss in the
structure into account [last term in the right-hand side of
Eq.(\ref{Kpropag})].

In fact, the independent variable 
parameterizing the dispersion curve 
can 
be chosen 
either as the wave number or as the frequency of
electromagnetic wave. 
%
%
%
%
%
%
%
%
%
%
%
%
In what follows, to obtain 
the standard Hamiltonian form of equations, 
the vector $\mathbf{k}$ is used 
as a real-valued
independent parameter.
%
%
%
Respectively, 
the quantity $\hbar\omega _{q\mathbf{k}}$ is 
a quantized observable
assuming complex values ($q$ is the band index). The real part of
observable $\hbar\omega _{q\mathbf{k}}$ 
is the photon energy and 
the
imaginary
part yields 
the lifetime broadening due to the optical loss.



%
%

Maxwell' equations in a photonic crystal with periodically varying
noninertiality (\ref{3Dg}) are solved here by separating the fast
and
slow field oscillations
in, respectively, 
longitudinal ($z$-axis) and lateral ($x$-$y$
plane) directions:
%
%
%
%
%
%
%
%
\begin{equation}
\left[
\begin{array}{c}
\mathbf{E}_{q\mathbf{k}} \\
\mathbf{H}_{q\mathbf{k}}%
\end{array}%
\right]
=
e^{-i\omega _{q\mathbf{k}}t}f_{q\mathbf{k}}(z)\left[
\begin{array}{c}
\sqrt{Z}\mathbf{e}_{q\mathbf{k}}(\mathbf{r}_{\bot }) \\
\frac{1}{\sqrt{Z}}\mathbf{h}_{q\mathbf{k}}(\mathbf{r}_{\bot })%
\end{array}%
\right], \label{SolutionForm}
\end{equation}
where the index $q$ enumerates 
the photonic bands, the scalar function $f_{q\mathbf{k}}(z)$ and
the set of two vector functions
$\mathbf{e}_{q\mathbf{k}}(\mathbf{r}_\perp)$ and
$\mathbf{h}_{q\mathbf{k}}(\mathbf{r}_\perp)$ are, respectively,
the fast and slow oscillating
 Bloch wave components. 
In (\ref{SolutionForm}), the field amplitudes
$\mathbf{e}_{q\mathbf{k}}$ and $\mathbf{h}_{q\mathbf{k}}$
are normalized 
in vacuum.
%
As shown 
below, such separation of variables is valid
in conditions of the paraxial approximation
 and low contrast of photonic crystal lattice. 

%



As
in 
(\ref{EinFP_off_ax}), the longitudinal part of the
wave (\ref{SolutionForm}) assumes a general form 
\begin{eqnarray}
f_{q\mathbf{k}}(z)=e^{ik_{z}z}\frac{e^{  i\phi
_{q\mathbf{k}}\sum_{j}\left[ \theta (z-2jL)-\frac{1}{2}\right] -
\frac{1}{2L}i\phi _{q\mathbf{k}}z }}{\sqrt{2\pi }} \label{fmK(z)}
\end{eqnarray}%
where the first and the second terms are, respectively, the plane
wave envelope function and 
periodic Bloch function part. 
The complex parameter $ \phi _{q \mathbf{k}} $ defines 
the magnitude of periodic
variations
in the Bloch function amplitude and phase.
This parameter 
is of the order of $\ln R \sim \xi^2$. Therefore, within the
accuracy of the $\xi^2$-order terms, the wave (\ref{fmK(z)})
assumes the expansion:
\begin{equation}
f_{q\mathbf{k}}(z) =e^{ik_{z}z} \frac{ 1+\eta _{q\mathbf{%
k}}(z)}{\sqrt{2\pi }} , 
\label{fmk(zzz)}
\end{equation}%
where
\begin{equation}
\eta _{q\mathbf{k}}(z)\simeq
 i\phi _{q\mathbf{k}}\sum_{j}\Bigl[ \theta (z-2jL)-\frac{1}{2}\Bigr] -\frac{i\phi_{q\mathbf{k}}z}{2L}.%
 \label{fmk(z)_aprox}
\end{equation}%
%
%
The weak periodic 
modulation $\eta
_{q\mathbf{k}}(z)$
shows 
step-like variations at the positions of $\mathbf{g}$-layers and thus represents 
the effect of 
abrupt phase-amplitude variations
at each reflection of the
mirror.
%
%
Note that the $z$-period average values of
%
%
the
$2L$-periodic function $\eta_{q\mathbf{k}}$ (odd function)
and its derivative $\partial \eta _{q\mathbf{k}}/
\partial z$ (even function) are null
%
($ \frac{1}{2L}\int_{-L}^{L} \eta _{q\mathbf{k}} (z)dz {=}
\left\langle \eta _{q\mathbf{k}} \right\rangle_{2L} {\simeq} 0$
and
$\left\langle \partial \eta _{q\mathbf{k}}/ \partial z%
\right\rangle _{2L}{\simeq} 0$). By virtue of these properties,
the longitudinal and lateral Bloch function components of 
electromagnetic wave (\ref{SolutionForm}) can be analyzed
separately. 

Substituting 
(\ref{SolutionForm})-(\ref{fmk(z)_aprox}) in Maxwell' equations
for the curl of $\mathbf{E}$ and $\mathbf{H}$,
and taking the $z$-period average,
one obtains the equations for the slow 
component
propagating in the lateral ($x$-$y$ plane) direction:
\begin{equation}
\begin{split}
&\left[\left(ik_{z} {-}\frac{\ln R(x,y
)}{2L} \right)
 \mathbf{\hat{z}}+
 \bm{\nabla} _{\bot }
 \right ] \times \mathbf{e}_{q\mathbf{k}}
{=} i\frac{n}{c} \omega _{q\mathbf{k}} \mathbf{h}_{q\mathbf{k}},
 \\
&\left[ \left(ik_{z} {-}\frac{\ln R(x,y
)}{2L} \right) \mathbf{\hat{z}}+ \bm\nabla _{\bot } \right]
\times \mathbf{h }_{q\mathbf{k}} {=} {-}i\frac{n}{c} \omega
_{q\mathbf{k}} \mathbf{e}_{q\mathbf{k}}.
\end{split} \label{SysEqs1}
\end{equation}%
%
%
%
Then, by taking the difference between the $z$-averaged
equation in (\ref{SysEqs1}) and corresponding Maxwell'
equation,
one gets 
the
equations for the fast oscillations 
in the $z$-axis direction
%
%
%
\begin{equation}
\begin{split}
\left[ \frac{\partial\eta _{q\mathbf{k}}}{\partial z} \right.
-\ln R\sum_{j}\delta (z-2jL)+ \left. \frac{\ln R}{2L} \right] 
\mathbf{\hat{z}} \times \mathbf{e}_{q \mathbf{k}} {=} 0,
 \\
\left[ \frac{\partial\eta _{q\mathbf{k}}}{\partial z} \right.
- \ln R\sum_{j}\delta
(z-2jL)+\left. \frac{\ln R}{2L} 
\right] \mathbf{\hat{z}}  \times \mathbf{h }_{q\mathbf{k} }{=}0.
\end{split}
\label{FastSys1_1}
\end{equation}

Eqs.(\ref{SysEqs1}) can be converted 
into a Hamiltonian eigenproblem, 
provided that a photonic wave function
$\bm\psi_{q\mathbf{k}}(\mathbf{r}_\perp)$ is introduced
via 
gauge transformation of the fields $\mathbf e
_{q\mathbf{k}}(\mathbf{r}_\perp)$ and $\mathbf
h_{q\mathbf{k}}(\mathbf{r}_\perp)$.
%
%


Wave states of a photon are usually 
expressed in terms of a three-component spinor wave
function
\cite{LandauIV} (a photon spin is 
one).
The spinor 
indexes 
represent polarization state, 
while the spatial distribution of spinor 
components 
accounts
for 
the angular momentum.
Thus, formally, 
three
spinor components
(with indexes $s_z{=}0,\pm1$) have to be taken into account 
in Eqs.
(\ref{SysEqs1})-(\ref{FastSys1_1}).

However, 
a 
state with $s_z{=}0$ cannot be realized in free
space\cite{LandauIV} and 
only two spinor components 
($s_z{=}\pm1$) are independent. 
Thus, by virtue of the transversality of electromagnetic wave in
vacuum,
only 
two 
field components 
might have independent spatial
distributions. 
%
%
Obviously, the same holds for homogeneous dielectric media or
periodic photonic crystal structures. In the last case, the wave
transversality condition
is replaced by the coupling of the field 
components
via Bragg scattering effects.

Therefore, a photonic state 
is 
defined here in terms of two independent spinor components
forming thus a 
two-component vector wave function
$\bm\psi_{q\mathbf{k}}(x,y)$.
For paraxial wave (\ref{SolutionForm}), it is convenient to
choose 
the 
components $\psi_{q\mathbf{k}}^{(x)}(x,y)$ and
$\psi_{q\mathbf{k}}^{(y)}(x,y)$  of wave function
$\bm\psi_{q\mathbf{k}}(x,y)$ in the lateral ($x$-$y$ plane)
direction. 
%
%
%
A perturbation 
analysis (not shown here)
reveals 
that
within the accuracy of $\xi^2$-order terms, 
%
the wave function components  can be introduced in Eqs.(\ref{SysEqs1})-(\ref{FastSys1_1})
via 
operator relationships 
%
%
%
%
\begin{equation}
\mathbf{e}_{q\mathbf{k}}= \hat{\bm{\mathcal{E}}}
\cdot \bm{\psi}_{q\mathbf{k}}, \quad \mathbf{h }_{q\mathbf{k}}=
\hat{\bm{\mathcal{E}}} \cdot
 \left[\mathbf{\hat{z}}\times
\bm{\psi}_{q\mathbf{k}} \right]  , \label{TzEqs}
\end{equation}
%
where $\left( \mathbf{\hat{z}}
\bm{\psi}_{q\mathbf{k}} \right) =0$, 
$\hat{\bm{\mathcal{E}}}$
is the tensor operator
%
\begin{eqnarray}
\hat{\mathcal{E}}_{\alpha\beta}{=} \delta_{\alpha\beta}
{+}\frac{i\delta_{\alpha 3}}{k_z}\frac{\partial}{\partial
x_{\beta} }
{+} \frac{1}{2k^2_z}
\Bigl(
 \frac{\partial^2}{\partial x_\alpha
\partial x_\beta}{-}
\frac{\delta_{\alpha\beta}}{2}
\frac{\partial^2}{\partial x_\gamma
\partial x_\gamma}
\Bigr),
%
\label{OperatorE}
\end{eqnarray}
and 
twice repeated Greek indexes indicate summation over the
$x$, $y$ and $z$ components.
 An explicit form of expressions
(\ref{TzEqs})-(\ref{OperatorE}) reads 
\begin{equation}
\begin{split}
\mathbf{e}_{q\mathbf{k}} &{=}
\Bigl(1{-}\frac{\triangle_{\perp}
}{4k_{z}^{2}} \Bigr)\bm{\psi}_{q\mathbf{k}}{+}i\mathbf{\hat{z}}%
\frac{\left(\bm\nabla _{\bot }\bm{\psi}_{q\mathbf{k}}\right) }{k_{z}} {+}\frac{%
\bm\nabla_{\bot}\left( \bm\nabla _{\bot }
\bm{\psi}_{q\mathbf{k}}\right) }{%
2k_{z}^{2}},
\\
\mathbf{h}_{q\mathbf{k}} &{=}
\Bigl(1{-}\frac{\triangle_{\perp}}{4k_{z}^{2}} \Bigr)
\mathbf{\hat{z}}{\times} \bm{\psi}_{q\mathbf{k}}
+i\mathbf{\hat{z}}%
\frac{\left( \bm\nabla _{\bot }\left[\mathbf{\hat{z}}{\times}
\bm{\psi}_{q\mathbf{k}}\right]\right) }{k_{z}} \\
&{+}\frac{\bm\nabla _{\bot }\left( \bm\nabla _{\bot }
\left[\mathbf{\hat{z}}{\times}
\bm{\psi}_{q\mathbf{k}}\right]\right) }{2k_{z}^{2}}.
\end{split}
\label{Phi-Chi}
\end{equation}
%
%
%
Eqs.
(\ref{TzEqs})-(\ref{OperatorE}) are used here as a definition of
the gauge transformation introducing a wave function
$\bm{\psi}_{q\mathbf{k}}$.
This gauge transformation 
is validated in 
Eqs.~(\ref{H}) and (\ref{ORT1}). 



%
%
%

In 
the gauge 
(\ref{TzEqs})-(\ref{OperatorE}), the
electromagnetic Bloch wave (\ref{SolutionForm}) reads
\begin{equation}
\left[
\begin{matrix}
E_{q\mathbf{k}}^{(\alpha)} \\
H_{q\mathbf{k}}^{(\gamma)}
\end{matrix}
  \right]
 \hspace{-0.05in}
  {=}e^{ik_z z{-}i\omega t} \frac{1{+}\eta(z)}{\sqrt{2\pi}}
\hspace{-0.04in} \left[
\begin{matrix}
Z^{\frac12} \hat{\mathcal{E}}_{\alpha\beta}\\
Z^{-\frac12}e_{3\beta\alpha}\hat{\mathcal{E}}_{\gamma\alpha}
\end{matrix}
\right]\hspace{-0.04in} \psi_{q\mathbf{k}}^{(\beta)} (x{,}y){,}
\label{Gauge}
\end{equation}
where
 $e_{\alpha\beta\gamma}$ is the completely antisymmetric
unit tensor of Levi-Civita.
%
The functions $\bm{\psi}_{q\mathbf{k}}$ and
$\mathbf{\hat{z}}{\times} \bm{\psi}_{q\mathbf{k}}$ defines
the electric and magnetic fields of the main polarization
component.
%
%
The first-order terms (${\propto}\frac1{k_z} \frac{\partial
}{\partial x} {\sim}\xi$)
%
%
and the second-order terms (${\propto}\frac1{k^2_z}
\frac{\partial^2}{\partial x^{2}}{\sim}\xi^2$)
%
%
contribute in 
longitudinal and cross polarization components of the wave.
Eq.~(\ref{Gauge}) 
allows a 
complex polarization structure of inhomogeneous wave
to be 
taken 
by the model into account and is in agreement with the results
obtained
for Gaussian beams.\cite{Erikson94}

%
%
%
%
%
%
%
%
%

It must be noticed 
that 
$\bm{\psi}_{q\mathbf{k}}$  (and $\mathbf{\hat{z}}{\times}
\bm{\psi}_{q\mathbf{k}}$)
differs from the lateral ($x$-$y$ plane) component  of the field 
$\mathbf{e}_{q\mathbf{k}}$ ($\mathbf{h}_{q\mathbf{k}}$). 
However, the squared modulus 
$\left | \bm{\psi}_{q\mathbf{k}}( \mathbf{r}_{\bot}
)\right |^2$ characterizes 
the 
energy flux in the $z$-axis direction.
%
Thus, taking a $z$-period average [Fig.\ref{fig1}(d)] of the
Poynting vector
$\mathbf{S}{=}\frac c {4 \pi} 
{\rm{Re}} \left\langle \frac12 \mathbf E \times \mathbf H^{*}
\right \rangle _ {2L}$, 
one obtains
$S_z{=}\frac c {8\pi} \left| \bm{\psi}_{q\mathbf{k}} \right|^2 $.
In the paraxial approximation considered here, the energy
flux 
in the $z$-axis direction  
is thus defined 
by the main polarization component of the wave. The longitudinal
and cross polarization components 
contribute to the energy flux 
in the lateral 
direction.
%
%

%
As an example, consider a wave function
$\bm{\psi}_{\mathbf{k}}=\bm{\psi}_{0}e^{ik_{x}x+ik_{y}y}$ with
$\bm{\psi}_{0}$ being a constant vector in the $x$-$y$ plane. It
defines a plane wave propagating in the direction of $\mathbf{k}=(
k_{x},k_{y},k_{z} )$. The energy flux associated with the wave is
of the density $S_{z} {=} \frac{c}{8\pi} \left| \bm{\psi}_{0}
\right|
^2$ along the 
$z$ axis. 
%
The
approximation (\ref{Gauge}) 
%
takes into account the transversality condition for electric and
magnetic components of a plane wave.
%
%
%
%
Thus, for the case of $p$-polarized (TM) wave 
($\mathbf{k}_\perp {\times} \bm\psi_0 {=} 0 $), 
the
electric field has nonzero components 
${\rm{e}}_{0z}{=}{-}k_\perp/k_z\psi_{0}$ 
and 
${\rm{e}}_{0\bot}{=}(1-\frac14 {k_\perp^2}/{k_z^2})\psi_{0}$,
yielding
the field amplitude 
${\rm{e}}_0{=}(1{+}\frac14 {k_\perp^2}/{k_z^2})\psi_{0}$. The
magnetic field oscillates 
in the $xy$-plane with the 
amplitude ${\rm{h}}_0{=}(1{+}\frac14
{k_\perp^2}/{k_z^2})\psi_{0}$.
%
The energy flux in the direction of $\mathbf{k}$ is thus
$S=S_{0z}(1+\frac12 {k_\perp^2}/{k_z^2})$, in agreement with the 
vector calculus 
utilizing 
directional angle $\xi{=}k_\perp/k_z$ of the wave.
By
substituting $\mathbf{e}\rightarrow - \mathbf {h}$ and
$\mathbf {h}\rightarrow \mathbf {e}$, a similar agreement 
can be readily proved 
for the case of 
$s$-polarized (TE) wave ($\mathbf{k}_\perp \bm\psi_0 {=} 0 $).
Within the accuracy of the $\xi^2$-order terms, the inversion of
Eq.~(\ref{Phi-Chi})
is straightforward, yielding
the 
expressions for the wave function $\bm\psi_{q\mathbf{k}}$ in
terms of the fields $\mathbf{e}_{q\mathbf{k}}$ and
$\hat{\mathbf{z}}\times \mathbf{h}_{q\mathbf{k}}$: 
\begin{equation}
\begin{split}
&\hspace{0.08in}\bm{\psi}_{q\mathbf{k}} {=} %
\Bigl(
 1{+} \frac{\triangle _{\perp }
}{4k_{z}^{2}}
\Bigr)
 \mathbf{e}_{q\mathbf{k}}{-}i\mathbf{\hat{z}} \frac{(
\nabla _{\bot }\mathbf{e}_{q\mathbf{k}}) }{k_{z}}
{-}\frac{ \nabla _{\bot } ( \nabla _{\bot }
\mathbf{e}_{q\mathbf{k}} ) }{ 2k_{z}^{2}},
\\
&\hspace{-0.1in}\mathbf{\hat{z}}{\times} \bm{\psi}_{q\mathbf{k}}
{=}
\Bigl( 1 {+} \frac{\triangle _{\perp }
}{4k_{z}^{2}}
\Bigr) \mathbf{h}_{q\mathbf{k}}{-}i\mathbf{\hat{z}}%
\frac{( \nabla _{\bot }\mathbf{h}_{q\mathbf{k}}) }{k_{z}}
{-}\frac{%
\nabla _{\bot }( \nabla _{\bot }
\mathbf{h}_{q\mathbf{k}}) }{%
2k_{z}^{2}}.
\end{split}
\label{Phi-Chi_inv}
\end{equation}
%
%
%
These relationships can be represented in the tensor operator
form that reads
%
\begin{equation}
\psi_{q\mathbf{k}}^{(\beta)}=\hat{\mathcal{E}}^{-1}_{\beta\alpha}
{\rm{e}} _{q\mathbf{k}}^{(\alpha)},
\quad  \psi_{q\mathbf{k}}^{(\alpha)} = e_{3 \alpha \beta}
\hat{\mathcal{E}}^{-1}_{\beta\gamma}
{\rm{h}}_{q\mathbf{k}}^{(\gamma)},
\label{InvTzEqs}
\end{equation}
%
%
%
%
%
where\cite{Note1} $e_{3\alpha\beta} e_{3\gamma\beta} {=}
\delta_{\alpha\gamma}-\delta_{\alpha3}\delta_{\gamma3}$, and the
inverse operator $\hat{\bm{\mathcal{E}}}^{-1}$ is defined by
\begin{eqnarray}
\hat{\mathcal{E}}_{\beta\alpha}^{-1}{=} \delta_{\alpha\beta}
{-}\frac{i\delta_{ \beta 3}}{k_z}\frac{\partial}{\partial
x_{\alpha} }
{-} \frac{1}{2k^2_z}
\Bigl(
 \frac{\partial^2}{\partial x_\alpha
\partial x_\beta}{-}
\frac{\delta_{\alpha\beta}}{2}
\frac{\partial^2}{\partial x_\gamma
\partial x_\gamma}
\Bigr)
%
\label{InvOperatorE}
\end{eqnarray}
%
Thus, $\hat{\bm{\mathcal{E}}}$  is not the unitary operator and
$\hat{\bm{\mathcal{E}}}^{-1}\neq\hat{\bm{\mathcal{E}}}^{+}$.

The paraxial gauge transformation (\ref{Gauge}) converts 
Maxwell' equations for the curl of $\mathbf{E}$ and $\mathbf{H}$
into the same form of a 
Hamiltonian eigenproblem with respect to the 
photonic state wave function $\bm{\psi}_{q\mathbf{k}}=\left(
\begin{smallmatrix} \psi_{q\mathbf{k}}^{(x)} \\
\psi_{q\mathbf{k}}^{(y)} \end{smallmatrix}
\right)$. Thus, substituting expressions 
(\ref{TzEqs}) in 
(\ref{SysEqs1}) and applying, respectively,
the operators
$e_{3\alpha\beta}\hat{\mathcal{E}}^{-1}_{\beta\gamma}$
and
$\hat{\mathcal{E}}^{-1}_{\beta\alpha}$, 
in 
the first and second equations [in (\ref{SysEqs1})]
one obtains 
\begin{equation}
\left( \frac{m_{0}c^2}{n^2} +\frac{\mathbf{\hat{p}}_{\bot
}^{2}}{2m_{0}}+i\frac{c\hbar }{n}\frac{\ln R(x,y)}{2L}\right)
\bm{\psi}_{q\mathbf{k}}=\hbar \omega _{q\mathbf{k}}
\bm{\psi}_{q\mathbf{k}}  \label{H}
\end{equation}%
where $m_0{=}n\hbar k_z/c $ is the effective mass and
$\mathbf{\hat{p}}_{\bot }{=}{-}i\hbar \mathbf{\nabla }_{\bot }$ is
the momentum operator in the lateral ($xy$-plane) direction.
%
%
%
%
%
%
The paraxial gauge transformation (\ref{Gauge}) representing an arbitrary photonic state by 
a function
$\bm{\psi}_{q\mathbf{k}} (\mathbf{r}_\perp)$ 
is thus validated by the fact that Maxwell' equations for $\mathbf{E}$ and $\mathbf{H}$ 
take the same form in this paraxial gauge. 


%
%
The first term
in the Hamiltonian $\hat H$ [right-hand side of Eq.(\ref{H})]  
is 
associated with the paraxial propagation along the cavity $z$
axis and accounts for the 
dispersion 
of the longitudinal 
wave vector component in the dielectric material of the cavity
($m_0c^2/n^2{=}\hbar k_zc/n $).
%
The in-plane kinetic energy  (second term in $\hat H$) and
effective 
potential (third term) take into
account the
dispersion 
and 
Bragg scattering effects in the lateral 
direction 
due to 
periodic 
reflectivity pattern $R(x,y)$ of the cavity mirror.
The parameter $m_{0}$ can
thus be interpreted as the lateral
effective mass of a photon
in an empty lattice (in the case of $R{=}1$).

The non-stationary Schr\"{o}dinger equation for
$\bm{\psi}_{q\mathbf{k}}(t,\mathbf{r}_\perp)$
follows from 
(\ref{H}) by 
substitution $ i
\partial /
\partial t \rightarrow \omega $.
In the case of $R{=}1$, within the accuracy of the time variable, the 
Schr\"{o}dinger
equation
(\ref{H})
is analytically similar to the scalar paraxial wave equation.
However, unlike a scalar field amplitude 
in paraxial wave equation, the spinor function
$\bm{\psi}_{q\mathbf{k}}$ in Eq.~(\ref{H}) cannot be associated
directly with any of the six components of electromagnetic field.
The same remark applies to comparison between the  Eq.(\ref{H})
and the scalar 2D Helmholtz equation for
microcavities.\cite{Hadley90}

The 
periodic 
potential $U(x,y){=}i\frac{c\hbar }{2nL}\ln R(x,y)$ implies that
$\bm{\psi}_{q\mathbf{k}}$ is a Bloch
wave\cite{Boiko04} composed of plane wave envelope function and
periodic crystal part
%
\begin{eqnarray}
\bm{\psi}_{q\mathbf{k}}(\mathbf{r}_\bot) &=& e^{i\mathbf{k}_\bot
\mathbf{r}_\bot}\mathbf{u}_{q\mathbf{k}}(\mathbf{r}_\bot),
\label{wavefunc}
\end{eqnarray}
where spinorial and angular parts of periodic function
$\mathbf{u}_{q\mathbf{k}}(\mathbf{r}_\bot)$
allow the impact of photonic lattice
symmetry to be taken into account 
in the analysis of
electromagnetic field behaviour 
under the lattice rotations. 
Since the
polarization anisotropy of mirror reflectivity 
$R(x,y)$ is much smaller than the order of effects accounted for
in the paraxial approximation
[see Eq.(\ref{reflection})], the spin-orbit coupling term is
neglected
in Eq.(\ref{H}). 
All eigen states of the Hamiltonian (\ref{H}) are thus doubly
degenerate by polarization. As shown in
Ref.~\onlinecite{Boiko07}, this degeneracy 
can be
removed by a symmetry breaking effects 
in 
nonpermanent gravitational field, when 
the photonic crystal rotates along the cavity $z$ axis.



Finally, the Hamiltonian $\hat H$
in Eq.(\ref{H}) is non-Hermitian 
($\hat H^+{\neq} \hat H$). 
The general properties of non-Hermitian Hamiltonians were
extensively studied in the past,
yielding the conditions for 
a discrete 
real-valued spectrum of
eigensolutions.\cite{Bender98,Mostafazadeh02} In Ref.
\onlinecite{Faisal81}, the interested reader can find a
comparison between the eigenproblems of Hermitian and
non-Hermitian Hamiltonians. In the case discussed here
[Eq.(\ref{H})], the most important results of these studies are
related to 
the spectrum of $\hat H$ and 
orthogonality condition of its eigenfunctions.

%
%
%
%
%
%
%
%
%
%


The Hamiltonian (\ref{H}) does not exhibit a pseudo-Hermiticity,
as opposed 
to a class of $PT$-symmetric Hamiltonians, which are invariant
under the time
reversal  
followed by coordinate inversion. 
%
%
Therefore, the spectrum of $\hat H$ assumes non-paired complex
eigenvalues $\hbar \omega_{q\mathbf{k}}$. Their complex conjugates
$\hbar \omega^{*}_{q\mathbf{k}}$ can only be obtained 
in the spectrum of the adjoint operator $\hat H^{+}$. The real
part of eigenvalues ${\rm{Re}} (\hbar \omega _{q\mathbf{k}})$ is
the photon energy and the imaginary part is the energy broadening
due to a finite
lifetime of photons in the cavity,
%
${\rm{Im}} (\hbar \omega
_{q\mathbf{k}})=\hbar/2\tau_{q\mathbf{k}}$.

%
%
%
%

%
%

%
%
%

The orthogonality of eigensolutions 
can be established using the 
biorthonormal set of functions\cite{Morse53} consisting of
%
the concomitant  partners $\bm{\psi}_{q\mathbf{k}}$ and
$\bm{\tilde\psi}_{{q}\mathbf{k}}$ of
associated with the eigenproblems
\begin{equation}
\hat{H}\left| \bm{\psi}_{q\mathbf{k}}\right){=}\hbar \omega
_{q\mathbf{k}} \left| \bm{\psi}_{q\mathbf{k}}\right) \label{eigH}
\end{equation}
and
\begin{equation}
\hat{H}^+|
\bm{\tilde{\psi}}_{q^{\prime}\mathbf{k}^{\prime}}){=}\hbar \omega
_{q^{\prime}\mathbf{k}^{\prime}}^* |
\bm{\tilde{\psi}}_{q^{\prime}\mathbf{k}^{\prime}}),
\label{eigHdag}
\end{equation}
respectively. Taking the difference between the matrix elements of
Eq.(\ref{eigH}) and the Hermitian adjoint of
Eq.(\ref{eigHdag}), one obtains 
\begin{equation} (\hbar \omega _{q\mathbf{k}}-\hbar \omega
_{q^{\prime}\mathbf{k}^{\prime}} )(\tilde{\psi}_{q^{\prime
}\mathbf{k}^{\prime }}^{(\alpha)}| \psi_{q\mathbf{k}}^{(\alpha)}
) {=}0. \label{eigHdiff}
\end{equation}
This relationship 
evidences\cite{LandauIII} the
orthogonality of 
functions associated with different
eigenvalues.
In 
the case of Hamiltonian (\ref{H}), the spinorial structure of
wave functions
$\bm{\psi}_{q\mathbf{k}}$ and $\bm{\tilde{\psi}}_{q\mathbf{k}}$ as well as the plane wave envelopes 
and 
periodic Bloch function parts\cite{Luttinger} 
have to be taken
%
by the 
orthogonality relationship into account, yielding %
%
%
\begin{eqnarray}
\hspace{-0.1in}(\tilde{\psi}_{q^{\prime }\mathbf{k}^{\prime
}}^{(\alpha)}| \psi_{q\mathbf{k}}^{(\alpha)} ) {=} \int
(\bm{\tilde{\psi}}_{q^{\prime } \mathbf{k}^{\prime }}^{*}{\cdot}
\bm{\psi}_{q\mathbf{k}}) d^{2}\mathbf{r}_{\bot } {=}\delta \left(
\mathbf{k} _{\bot }-\mathbf{k}_{\bot }^{\prime }\right) \delta
_{q^{\prime }q}. \label{VmkNorm}
\end{eqnarray}%
Here, the integration runs over the entire $x$-$y$ plane and $
(\bm{\tilde{\psi}}_{q^{\prime } \mathbf{k}^{\prime }}^{*}{\cdot}
\bm{\psi}_{q\mathbf{k}}){=}\tilde{\psi}_{q^{\prime
}\mathbf{k}}^{(\alpha)}(x,y)^{*}\psi_{q\mathbf{k}}^{(\alpha)}(x,y)$
is the scalar product evaluated at a point
$\mathbf{r}_\perp{=}(x,y)$. For photonic states at the same
point of
 the 2D Brillouin zone ($\mathbf
{k}^{\prime}_{\bot}{=}\mathbf k_{\bot} $),
 the integral over the entire crystal (\ref{VmkNorm}) can be reduced 
 to a
 lattice-cell integral\cite{Luttinger} 
\begin{equation}
 \langle \tilde{\psi}_{q^{\prime
}\mathbf{k}}^{(\alpha)}|
\psi_{q\mathbf{k}}^{(\alpha)} \rangle = \frac{\left( 2\pi \right)
^{2}}{ \Omega _{\bot }}\int_{\text{cell}}
(\bm{\tilde{\psi}}_{q^{\prime }\mathbf{k}}^{*} {\cdot}
\bm{\psi}_{q\mathbf{k}}) d^{2}\mathbf{r}_{\bot } = \delta
_{q^{\prime }q}, \label{VmkNorm2}
\end{equation}%
where $\Omega _{\bot }$ is the 
lattice cell area.
%
%
%
%
%
%
%
%
For a state
$|\bm{\psi}_{q\mathbf{k}}\rangle$,
 the probability distribution function
for the coordinates is thus given by 
the product
$\bm{\tilde{\psi}}_{q\mathbf{k}}^{*}{\cdot}\bm{\psi}_{q\mathbf{k}}$,
as opposed to the usual expression $|\bm{\psi}_{q\mathbf{k}}|^2$.
Indeed
the probability to find a photon at a point $\mathbf
r_{\perp}^{\prime}$ 
is $\langle
\tilde{\psi}_{q\mathbf{k}}^{(\alpha)}|\delta(\hat{\mathbf
r}_{\perp}{-}\mathbf r_{\perp}^{\prime})|
\psi_{q\mathbf{k}}^{(\alpha)} \rangle{=}
\bm{\tilde{\psi}}_{q\mathbf{k}}^{*} (\mathbf r_{\perp}^{\prime})
{\cdot}\bm{\psi}_{q\mathbf{k}}(\mathbf r_{\perp}^{\prime})$. In
the Schr\"{o}dinger picture, this expression allows
the stationary probability distribution to be associated with an
eigen state. By contrast, the distribution
$|\bm{\psi}_{q\mathbf{k}}|^2$
varies in time as $\propto \exp{(-t/\tau_{q\mathbf{k}})}$.
This behaviour is expected by the photon lifetime considerations,
since 
 $|\bm{\psi}_{q\mathbf{k}}|^2$ defines the 
the density of the energy flux along the cavity $z$-axis. Note
that in the case of Hermitian Hamiltonian [$\text{Im} (U){=}0$],
both
distributions are 
stationary and indistinguishable.
%
%
%
%
%
%
%
%


Substituting  a Bloch wave (\ref{wavefunc}) for the
$\bm{\psi}_{q\mathbf{k}}$ and using the expression
$\bm{\tilde{\psi}}_{q\mathbf{k}} = e^{i\mathbf{k}_\bot^{*}
\mathbf{r}_\bot}\mathbf{\tilde{u}}_{q\mathbf{k}}(\mathbf{r}_\bot)$
for its concomitant partner, we obtain the orthogonality
relationship for the periodic Bloch functions\cite{Note2}
%
%
\begin{equation}
 \langle \tilde{u}_{q^{\prime
}\mathbf{k}}^{(\alpha)}|
u_{q\mathbf{k}}^{(\alpha)} \rangle = \frac{\left( 2\pi \right)
^{2}}{ \Omega _{\bot }}\int_{\text{cell}}
(\bm{\tilde{u}}_{q^{\prime }\mathbf{k}}^{*} {\cdot}
\bm{u}_{q\mathbf{k}}) d^{2}\mathbf{r}_{\bot } = \delta _{q^{\prime
}q}. \label{VmkNormBloch}
\end{equation}%

The orthogonality of 
partners $\bm{\psi}_{q\mathbf{k}}$ and
$\bm{\tilde{\psi}}_{q^{\prime}\mathbf{k}^{\prime}}$ 
leads to the orthogonality relationship for the lateral
components of the electric and magnetic fields
(\ref{TzEqs}).
Since the gauge transformation operator $\bm{\hat{\mathcal{E}}}$
[Eq.~(\ref{OperatorE})] is independent of the material equations,
the electromagnetic field
associated
with the 
partner $\bm{\tilde{\psi}}_{q\mathbf{k}}$
is defined 
by 
the 
relationships
\begin{equation}
\mathbf{\tilde{e}}_{q\mathbf{k}}{=} \bm{\hat{\mathcal{E}}}
{\cdot} \bm{\tilde{\psi}}_{q\mathbf{k}}, \quad \mathbf{\tilde {h
}}_{q\mathbf{k}}{=} \bm{\hat{\mathcal{E}}} {\cdot}
 [\mathbf{\hat{z}}{\times}
\bm{\tilde{\psi}}_{q\mathbf{k}} ].
\end{equation}
The scalar product in the integrand of Eq.~(\ref{VmkNorm}) can be
then
expressed 
in terms of the electric and magnetic fields
%
\begin{equation}
\begin{split}
&{\int}(\bm{\tilde{\psi}}^{*}_{q^\prime\mathbf{k}^\prime}
{\cdot}\bm{\psi}_{q\mathbf{k}}) d^2 \mathbf{r}_\perp \\
&\quad\quad\quad
{=}{-}{\int}\hat{\mathbf{z}}{\cdot}\Bigl(\hat{\bm{\mathcal{E}}}^{-1}\tilde{\mathbf{h
}}_{q^\prime\mathbf{k}^\prime}\Bigr)^{*}{\times}
\Bigl(\hat{\bm{\mathcal{E}}}^{-1}\mathbf{e }_{q\mathbf{k}}\Bigr)
 d^2 \mathbf{r}_\perp.
\end{split}
\label{NormhEepsilon_Inv}
\end{equation}
Substituting 
Eq.~(\ref{InvOperatorE}) for $ \hat{\bm{\mathcal{E}}}^{-1}$ and
using the relationship $\hat{\mathbf{z}}
{\times} (\nabla _{\bot }( \nabla _{\bot }
\mathbf{\tilde{h}}^{*}_{q^\prime\mathbf{k}^\prime}))
=
\triangle
_{\perp } [\hat{\mathbf{z}} {\times}
\mathbf{\tilde{h}}^{*}_{q^\prime\mathbf{k}^\prime}]
-
\nabla _{\bot }( \nabla _{\bot }
[\hat{\mathbf{z}} {\times}
\mathbf{\tilde{h}}^{*}_{q^\prime\mathbf{k}^\prime}])$, one
obtains that 
\begin{equation}
\begin{split}
&{\int}(\bm{\tilde{\psi}}^{*}_{q^\prime\mathbf{k}^\prime}
{\cdot}\bm{\psi}_{q\mathbf{k}}) d^2 \mathbf{r}_\perp
{=}{-}\int\hat{\mathbf{z}}{\cdot}[\tilde{\mathbf{h
}}^{*}_{q^\prime\mathbf{k}^\prime} {\times} \mathbf{e
}_{q\mathbf{k}}
 ]d^2 \mathbf{r}_\perp\\
& \quad \quad \quad \quad {-} \frac 1 {4k_{z}^{2}} \int
\Bigl(
\bm{\nabla}_{\bot } {\cdot}
\Bigl\{
([\hat{\mathbf{z}} {\times}
 \tilde{\mathbf{h
}}^{*}_{q^\prime\mathbf{k}^\prime}
 ])^{(\alpha)}
\bm{\nabla}_{\perp } \text{e}^{(\alpha)}_{q\mathbf{k}}
\\
&\quad \quad \quad \quad {-} \text{e}^{(\alpha)}_{q\mathbf{k}}
\bm{\nabla}_{\perp } ([
 \hat{\mathbf{z}}
{\times}
\mathbf{\tilde{h}}^{*}_{q^\prime\mathbf{k}^\prime}])^{(\alpha)}
\Bigr\} \Bigr)d^2 \mathbf{r}_\bot
 \\
&\quad \quad \quad \quad {+} \frac 1 {2k_{z}^{2}} \int
\Bigl( \bm{\nabla} _{\bot } {\cdot}
\Bigl\{ [\hat{\mathbf{z}}
{\times}
 \tilde{\mathbf{h
}}^{*}_{q^\prime\mathbf{k}^\prime}
 ]
( \bm{\nabla} _{\bot } {\cdot}
\mathbf{e}_{q\mathbf{k}} )\\
&\quad \quad \quad \quad
 {-} \mathbf{e }_{q\mathbf{k}} ( \bm{\nabla} _{\bot }{\cdot}
[
 \hat{\mathbf{z}}
{\times}\mathbf{\tilde{h}}^{*}_{q^\prime\mathbf{k}^\prime}])
\Bigr\} \Bigr ) d^2 \mathbf{r}_\perp,
 \\
\end{split}
\label{Interim_ORT1_eh}
\end{equation}
where twice repeated index $\alpha$ indicates summation over the
$x$, $y$ and $z$ vector field components. In the right-hand side
of Eq.(\ref{Interim_ORT1_eh}), the integrands in the second and
third terms
%
%
are of the form $(\bm{\nabla}_{\bot} {\cdot} \mathbf{F})$ with
$\mathbf{F}$ being a 
vector field given by one of the expressions in curly brackets.
%
%
As
in 
Eqs.(\ref{VmkNorm})-(\ref{VmkNorm2}), the plane wave envelope
function and periodic Bloch part\cite{Luttinger} 
have to be
accounted for in the electric and magnetic field components
contributing to $\mathbf{F}$.
%
%
It follows that $\mathbf{F}$ can 
be represented as a product $\mathbf{F}{=}
e^{i(\mathbf{k}_\perp{-}\mathbf{k}^\prime_\perp)\mathbf{r}_\perp}\mathbf
{f}(\mathbf{r}_\perp)$ with $\mathbf {f}(\mathbf{r}_\perp)$ being
a cell-periodic function.
%
%
%
%
%
%

Integration 
over the 
$x$-$y$ plane can be then accomplished
%
%
by using the standard methods of the crystal field
theory,\cite{Luttinger}
%
%
yielding
%
%
$\int (\bm{\nabla}_{\bot} {\cdot} \mathbf{F}) d^2
\mathbf{r}_\perp=
\delta(\mathbf{k}_\perp{-}\mathbf{k}^\prime_\perp) \frac{4
\pi^2}{\Omega_\perp} \int_{\text{cell}} (\bm{\nabla}_{\bot}
{\cdot} \mathbf{f})  d^2 \mathbf{r}_\perp $.
Stokes' theorem
allows
the last integral
over the lattice cell to be transformed into the 
contour integral along the cell boundaries yielding
\begin{equation}
\int_\text{cell} \!\!\! (\bm{\nabla}_{\bot} {\cdot} \mathbf{f})
d^2 \mathbf{r}_\perp{=}\!\int_\text{cell} \!\!\!
\bm{\nabla}_{\bot} {\times } [ \mathbf{\hat{z}} {\times}
\mathbf{f}] {\cdot} d \bm{S}_\perp{=} \! \oint_{\partial
\{\text{cell}\}}\! \!\!\! \hat{\mathbf{z}}
 {\times} \mathbf{f}  {\cdot} d \mathbf{l}_\bot
 \nonumber
\end{equation}
where $d \mathbf{S}_\bot$ and $d \mathbf{l}_\bot$ are,
respectively,
the elements of the Wigner-Seitz cell and its boundary,
and $\partial \{\text{cell}\}$ is 
the
counterclockwise oriented contour in the $x$-$y$ plane.
%

The symmetry of the Wigner-Seitz cell assumes that for each point
at the boundary, 
one can
put in correspondence another point at the opposite boundary
such that the two points are related by a primitive
lattice translation. The contributions from such points in the
integral $ \oint_{\partial \{\text{cell}\}}  \hat{\mathbf{z}}
 \times \mathbf{f}  \cdot d \mathbf{l}_\bot $ cancel
out each other
since at these points, the periodic function
$\mathbf{f}(\mathbf{r}_\perp)$ takes the same values, while the
contour elements $d \mathbf{l}_\perp$ are oriented in opposite
directions.
%

It 
 follows that the second and third integrals in the
right-hand side of Eq.(\ref{Interim_ORT1_eh}) vanish.
Therefore, 
by virtue of (\ref{VmkNorm}),  the orthogonality relationship for
the slowly varying
components (\ref{TzEqs}) of the 
field reads
%
%
\begin{equation}
\int \mathbf{\hat{z}} {\cdot} [\mathbf{\tilde{h}}_{q^{\prime
}\mathbf{k}^{\prime }}^{*} {\times} \mathbf{e }_{q\mathbf{k}}]
d^{2}\mathbf{r}_{\bot } {=} - \delta \left( \mathbf{k}_{\bot }-
\mathbf{k}_{\bot }^{\prime
}\right) \delta _{q^{\prime }q},  
\label{ORT1_eh}
\end{equation}
where the integration runs over the 
$x$-$y$ plane.
%
%
%
%
%
Thus, within the accuracy of the paraxial approximation,
the orthogonality relationship
between the wave states 
from different bands [with indexes ${q}$ and $q^{\prime}$ in
Eq.(\ref{ORT1_eh})] is not influenced by
the fast 
longitudinal (along $z$ axis) oscillations (\ref{fmk(zzz)}) of
the electromagnetic
field. 
Using (\ref{SolutionForm}), 
separating the integrals for longitudinal envelope functions and
periodic parts  [as in Eq.~(\ref{VmkNormBloch})],
and noting 
that 
$\langle\eta_{q\mathbf{k}}\rangle_{2L}{=}0$
with
accuracy
${\sim}\xi^2$, 
one 
obtains
the orthogonality relationship for
electromagnetic 
waves
that reads
\begin{equation}
\int \mathbf{\hat{z}} {\cdot}
[\mathbf{\tilde{H}}_{q^{\prime }\mathbf{k}^{\prime }}^{*}
{\times} \mathbf{E}_{q\mathbf{k}}]
d^{3}\mathbf{r}_{\bot } {=} {-} 
\delta (k_{z}{-} k_{z}^{\prime })\delta \left( \mathbf{k}_{\bot
}{-} \mathbf{k}_{\bot }^{\prime }\right) \delta _{q^{\prime }q},
\label{ORT1}
\end{equation}
where 
integration is 
over the entire cavity-unfolded structure.
%
%
%
%
%
Note that relationship (\ref{ORT1})
is 
in agreement with 
the waveguiding theories.\cite{Kogelnik75,Slater46}

With the help of orthogonality relationship (\ref{VmkNorm2}), the
solution of equations for the fast longitudinal oscillations of
the field 
[Eq.~(\ref{FastSys1_1})] is straightforward. As in the case of
Eq.~(\ref{H}), it follows
by applying the 
gauge transformation (\ref{Phi-Chi}) and 
the operators
${-}{\bm{\hat{z}}}{\times}\bm{\hat{\mathcal{E}}}^{-1}$ and
$\hat{\bm{\mathcal{E}}}^{-1}$ in the first and second equations
in (\ref{FastSys1_1}).
%
Within the accuracy of $\xi^2$-order terms,
both equations 
yield 
the same form of the master equation for the periodic part $\eta
_{q\mathbf{k}}(z)$ of the fast longitudinal wave component
\begin{equation}
\frac{\partial \eta _{q\mathbf{k}}(z)}{\partial z}=
  i\phi_{q\mathbf{k} }\sum_{j}\delta
(z-2jL)-\frac{i\phi_{q\mathbf{k} }}{2L}, \label{EqLongit}
\end{equation}%
where 
$\phi_{q\mathbf{k}}{\sim}\xi^2$ (or less) is the diagonal matrix
element $\phi_{q\mathbf{k}} {=} {-}i\langle
\tilde{\psi}_{q\mathbf{k}}^{(\alpha)}| \ln R |
{\psi}_{q\mathbf{k}}^{(\alpha)} \rangle$.
%
%
The off-diagonal 
elements have 
smaller 
magnitudes and, within the accuracy 
${\sim}\xi^2$,
their contribution cannot be accounted for in the right-hand side
of Eq.(\ref{EqLongit}).
In this case, the left-hand side of Eq.~(\ref{EqLongit})
is also null 
($  \langle
\tilde{\psi}_{q^{\prime}\mathbf{k}}^{(\alpha)}|
 {\psi}_{q\mathbf{k}}^{(\alpha)} \rangle \partial
\eta _{q\mathbf{k}}/\partial z {\equiv} 0 $ for $q^{\prime}\neq
q$).

The integral of Eq.(\ref{EqLongit}) 
is
%
the expression
given in (\ref{fmk(z)_aprox}) but now with the
parameter $\phi_{q\mathbf{k}}$ 
expressed 
as the diagonal matrix element of effective crystal potential.
%
%
This verifies 
the separation of variables in the paraxial gauge transformation
(\ref{Gauge}).

The matrix elements $\phi_{q\mathbf{k}}$
provide a way for more accurate formulation 
of  the paraxial approximation conditions in the Hamiltonian
(\ref{H}).
%
%
%
%
%
%
%
%
Thus, the effective crystal potential
in 
(\ref{H}) has %
been assumed so far to be of the relative order $\xi^2$,
that is
$ \frac{c\hbar }{n}\frac{|\ln R|}{2L} {\lesssim} \xi^2
\frac{m_{0}c^2}{n^2}$. Using 
the roundtrip self-repetition condition in a $\lambda$-cavity
$k_z{=}2\pi/L$,
the  
 condition of low-contrast patterning 
can be represented in the form
\begin{equation}
| \ln R(x,y) | \lesssim 4\pi \xi^2.  \label{dR_valid_0}
\end{equation}
%
%
However, as evidenced by Eq.~(\ref{EqLongit}),
it
is more important that 
the effective 
potential variations are small in average,
at the scale of the crystal lattice cell.
For a particular photonic band, the last condition can be
expressed 
by evaluating
 the diagonal matrix element in (\ref{dR_valid_0}), yielding 
%
%
\begin{equation}
|\langle \tilde{\psi}_{q\mathbf{k}}^{(\alpha)}| \ln R(x,y) |
{\psi}_{q\mathbf{k}}^{(\alpha)} \rangle | \alt 4\pi \xi^2,
\label{dR_MatrEq_Valid}
\end{equation}
where the 
matrix element 
is the same as 
$|\phi_{q\mathbf{k}}|$ 
in Eq.~(\ref{EqLongit}).

The condition 
(\ref{dR_MatrEq_Valid}) 
verifies 
the 
paraxial
approximation (\ref{H}) within a particular set of
photonic bands.
%
For 
structures with 
simple lattice cell topology 
(like in Fig.~\ref{fig1}), this condition 
is less restrictive.
%
%
For such structures, further simplification is possible in the
most important case of low-order photonic bands, which are
typically characterized by relatively smooth wave functions.
%
Introducing 
the reflectivity contrast parameter 
$\delta R$ as a measure of the maximum variations in 
$\ln R(x,y)$, 
the intraband matrix elements (\ref{dR_MatrEq_Valid}) can be then
estimated as $|\langle \tilde{\psi}_{q\mathbf{k}}^{(\alpha)}| \ln
R | {\psi}_{q\mathbf{k}}^{(\alpha)} \rangle | {\sim} FF | \delta
R |$ for the lattices with small fill factor $FF{<}\frac 12$
(\emph{e.g.}, for the lattices with small-size pixels in the case of %
structures shown in Fig.\ref{fig1}) .
%
%
In the opposite case of large fill factor $FF{>}\frac 12$
(large-sized pixels in Fig.\ref{fig1}), the
matrix elements 
are 
$ {\sim}
(1-FF) | \delta R |$. 
%

It follows that the 
paraxial Hamiltonian (\ref{H})
is valid 
for the 
structures with reflectivity patterning contrast 
\begin{equation}
\begin{split}
|\delta R |&\alt \xi^2 4 \pi /FF \quad \quad \quad (FF{\leq}1/2), \\
&\alt \xi^2 4 \pi /(1{-}FF) \quad (FF{>}1/2).
\end{split}\label{dR_valid}
\end{equation}
In most practical
cases, 
optical microcavities satisfy 
this condition.

\section{Dielectric lattices 
defined by refractive index
variations}
\label{SecIndexPatterned}

\begin{figure}[tbp]
\includegraphics{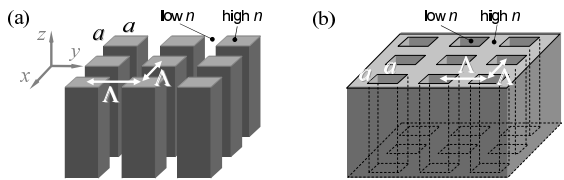}
\caption{
Schematic illustration of 
the paraxial photonic crystal structures defined by periodic
variations of refractive index.
(a) 
Arrays of coupled index-guided 
 microcavities
(\emph{e.g.}, etched VCSEL arrays\cite{Gourley91}) or parallel
waveguides (like in the cladding region of photonic band gap
fibers\cite{Argyros05}).
(b) Complementary, antiguided-array structures  
of microcavities defined by negative-index step (\emph{e.g.},
antiguided VCSEL arrays\cite{Mawst03}) or
low-index core antiguides (like in photonic liquid crystal
fibers\cite{Wolinski06}).
  $\Lambda$ is the lattice pitch, $a$ is the pixel size.}
\label{fig2}
\end{figure}


The 
approach developed in Sec.\ref{SecMirrorPatterned} 
is 
focused 
on 
lattices of optical resonators realized by mirror reflectivity
patterning 
 (Fig.\ref{fig1}).
For completeness of the Hamiltonian (\ref{H}), it has to be
extended to similar
lattices 
of coupled microcavities or parallel waveguides defined 
by periodic variations of the dielectric constant
(Fig.\ref{fig2}). Examples of such coupled microcavities
with positive [Fig.\ref{fig2}(a)] or negative [Fig.\ref{fig2}(b)]
refractive index contrast
 are, respectively, etched VCSEL arrays\cite{Gourley91} or matrices
of antiguided VCSEL resonators 
fabricated using a two-step organometallic chemical vapor
deposition (OMCVD) growth.\cite{Mawst03} As shown below, the
approach based on Hamiltonian formalism (\ref{H}) applies equally
well to arrays of parallel waveguides\cite{Shteeman07}
[Fig.\ref{fig2}(a)] and low-index-core antiguides
[Fig.\ref{fig2}(b)].
These photonic crystal materials are 
employed in the cladding of photonic band gap
fibers\cite{Argyros05} and photonic liquid crystal
fibers.\cite{Wolinski06}

In this section, the Hamiltonian (\ref{H}) is thus extended to
almost entire
subclass of 2D paraxial photonic crystal 
structures 
characterized by the 
light propagation mostly normal to the periodic crystal plane. As
it can be
expected, 
the Hamiltonian for such 
dielectric lattices [see Eq.(\ref{H_wgd})] is analytically
similar to the scalar paraxial wave equation 
but differs from it by the use of two-component spinor wave
functions for photons.

Note that a general form of the Hamiltonian for dielectric
lattices
can be obtained directly from Eq.~(\ref{H}), assuming that
$R{=}1$ 
and  introducing the refractive index variations
$n(x,y)^{-1} {-} \langle n^{-1} \rangle$
and 
effective mass $m_0{=} \hbar k_z/c\langle n^{-1} \rangle
$ in 
the first term of 
Eq.(\ref{H}). These 
yield the effective periodic 
potential of
$U
{=}m_{0}c^2 \langle \frac1n \rangle ( \frac 1{n(x,y)} {-} \langle
\frac1n \rangle ) $.
However, such
calculus does not
reveal 
the maximum refractive index contrast $\delta n $ satisfying %
the conditions of
paraxial Hamiltonian approximation.
As shown below, the allowed refractive index variations assume
the condition (\ref{dn_MatrEq_Valid}), which for low-order
photonic bands, can be represented in the form
%
%
%
%
%
%
%
%
\begin{equation}
\begin{split}
\Bigl |\delta n \Bigl \langle \frac 1n \Bigr \rangle \Bigr| & \alt \xi /FF  \quad \quad \quad (FF\leq 1/2),\\
& \alt \xi /(1{-}FF) \quad (FF > 1/2),
\end{split}\label{dn_valid}
\end{equation}
where 
$FF$ is the lattice cell fill factor, $\Lambda$ is the lattice
pitch, $\lambda$ is the wavelength (in vacuum) and $\xi =\langle
n^{-1} \rangle\lambda/2 \Lambda $ is the order parameter
($\xi{\ll} 1$)
corresponding 
to a propagation angle
of 
state 
at the boundary of the first (2D) Brillouin zone. 
%
%
%
%
Unlike the second-order 
variations $\delta R $ in  
reflectivity-patterned
structures, the refractive index
variations $\delta n $
can thus be
treated by the model 
as 
first-order perturbations 
 [compare Eqs.(\ref{dR_valid}) and (\ref{dn_valid})]. As discussed in Sec.
\ref{ResultsAndDiscussion}, this allows the paraxial Hamiltonian
to be applied 
to some of the structures exhibiting high contrast of refractive
index variations ($\delta n {\sim} 1$), as 
in the case of
holey photonic crystal fibers\cite{Russell03} or arrays of
micropillars.\cite{Bayer99}
%
%
%

%

%
%

To analyze separately the impact of 
refractive index
variations
on photonic band structure, we assume that $\mathbf g{=}0 $ in
constitutive 
equations (\ref{MatEqED}). 
 In the case of coupled microcavities, this suggests that the
uniform loss and gain distributions are neglected and
the reflectivity of the output coupling
mirror is $R{=}1$.
As in the case of mirror-patterned structures, 
the vertical cavity
composition is accounted for by using an effective (complex)
refractive index profile\cite{Hadley90} $n(x,y)$. The Bragg
scattering effects 
set in, conditioned by the periodic
pattern $\varepsilon(x,y){=}n(x,y)^2/\mu$.
For the
structures considered here,
the magnetic permeability $\mu$ is assumed
 constant ($\mu=1$ in the optical spectrum
range). Thus,
in accordance with
 the relationships $\varepsilon(x,y) = n(x,y)/Z(x,y) $ and $\mu =
n(x,y) Z(x,y)$, both the 
refractive index $n(x,y)$ and the impedance $Z(x,y)$ exhibit
periodic variations.

%
%

The cavity-unfolded representation 
(Sec.~\ref{SecMirrorPatterned}) can be applied to the matrices of
dielectric microcavities
as well. In this particular case, it effectively translates the
cavity into a structure that exhibits a translational symmetry
along cavity $z$-axis, allowing thus the correspondence between an
array of microcavities and equivalent structure of parallel
dielectric waveguides. Electromagnetic waves propagating in such
waveguiding
structures (in the $z$-axis direction) 
can be expressed using the paraxial wave approximation
(\ref{SolutionForm}) with separated fast (longitudinal) and slow
(lateral) wave oscillations.
%
%

Since $R{=}1$,
an electromagnetic wave propagating
in 
a cavity-unfolded structure
does not show sudden amplitude variations at 
$\mathbf{g}$-layers 
[see Sec.~\ref{SecMatEq}]. Therefore,
as in 
the arrays of parallel waveguides 
as in 
the cavity-unfolded lattices of dielectric microcavities, the fast
longitudinal 
component of the wave 
is a uniform plane wave 
showing no periodic 
modulation 
in the
$z$-axis direction [see Eq.(\ref{fmk(zzz)})]: 
\begin{equation}
\eta_{q\mathbf{k}}(z)=0.
\label{FastSys1_1__nZ}
\end{equation}
The electric and magnetic field components then read
\begin{equation}
\left[
\begin{array}{c}
\mathbf{E}_{q\mathbf{k}} \\
\mathbf{H}_{q\mathbf{k}}%
\end{array}%
\right] =
e^{-i\omega _{q\mathbf{k}}t}\frac{ e^{ik_{z}z}}{\sqrt{2\pi
}}\left[
\begin{array}{c}
Z(\mathbf{r}_{\bot })^{1/2} \mathbf{e}_{q\mathbf{k}}(\mathbf{r}_{\bot }) \\
Z(\mathbf{r}_{\bot })^{-1/2}\mathbf{h}_{q\mathbf{k}}(\mathbf{r}_{\bot })%
\end{array}%
\right] \label{SolutionForm_dn}
\end{equation}
%
%
After substitution of (\ref{SolutionForm_dn}), Maxwell's equations
for the curl of $\mathbf{E}$ and $\mathbf{H}$
yields 
%
the equations for the slowly varying wave components 
in the lateral $xy$-plane direction:
\begin{equation}
\begin{split}
&\frac{c}{n(x,y)}\left[ik_{z}
 \mathbf{\hat{z}}{+}
 \bm{\nabla} _{\bot }
  {+} \frac{\bm{\nabla}_{\bot}( Z )}{2Z(x,y)}  \right] {\times} \mathbf{e}_{q\mathbf{k}}
= i \omega _{q\mathbf{k}} \mathbf{h}_{q\mathbf{k}}
 \\
&\frac{c}{n(x,y)}\left[ ik_{z}
\mathbf{\hat{z}}{+} \bm\nabla _{\bot }{-}
\frac{\bm{\nabla}_{\bot}( Z )}{2Z(x,y)}
 \right] {\times} \mathbf{h }_{q\mathbf{k}}
= {-}i \omega _{q\mathbf{k}} \mathbf{e}_{q\mathbf{k}}
\end{split} \label{SysEqs1_nZ}
\end{equation}%
In order to convert these equations
into the Hamiltonian eigenproblem, an appropriate gauge
transformation has to be applied for the fields
$\mathbf{e}_{q\mathbf{k}}$ and $\mathbf{h}_{q\mathbf{k}}$.
%
%
%
%
A 
general form of the gauge transformation, which 
takes into account the variations of $Z(x,y)$ and $n(x,y)=\mu/
Z(x,y)$ and, at the same time, encompasses the particular case of
a uniform dielectric medium (\ref{Gauge}), reads:
\begin{equation}
\mathbf{e}_{q\mathbf{k}}= Z^{a} \hat{\bm{\mathcal{E}}}
Z^{-a}\cdot \bm{\psi}_{q\mathbf{k}}, \quad \mathbf{h
}_{q\mathbf{k}}= Z^{-b}
\hat{\bm{\mathcal{E}}}Z^{b} \cdot
  \mathbf{\hat{z}}\times
\bm{\psi}_{q\mathbf{k}}, \label{TzEqs-dn}
\end{equation}
where, as 
in 
(\ref{Gauge}), $\bm{\psi}_{q\mathbf{k}} $ is a two-component
spinor wave function, the tensor operator
$\hat{\bm{\mathcal{E}}}$ is defined in (\ref{OperatorE}) and
the  parameters $a$ and $b$ will be 
adjusted 
to fit 
the spinor transversality condition
$\left(
\mathbf{\hat{z}}
\bm{\psi}_{q\mathbf{k}} \right) =0$.
%
%
%
%

The corresponding inverse transformation reads
\begin{equation}
\bm{\psi}_{q\mathbf{k}}{=}Z^{a}\hat{\bm{\mathcal{E}}}^{-1} Z^{-a}
{\cdot} \mathbf{e}_{q\mathbf{k}}, \quad \mathbf{\hat{z}}{\times}
\bm{\psi}_{q\mathbf{k}}{=}Z^{-b}\hat{\bm{\mathcal{E}}}^{-1}Z^{b}{\cdot}\mathbf{h
}_{q\mathbf{k}}, \label{InvTzEqs-dn}
\end{equation}
where $\hat{\bm{\mathcal{E}}}^{-1}$ is the tensor operator
introduced in Eq.(\ref{InvOperatorE}).
Note that
$1/n$ can be substituted for $Z$ in
Eqs.(\ref{TzEqs-dn})-(\ref{InvTzEqs-dn}) since for considered
here structures, the magnetic permeability $\mu$ is constant.

In these expressions, the spatial derivatives of $Z$ (or $1/n$)
should match the paraxial approximation conditions, which thus
limit the contrast of dielectric materials composing the 
lattice. Introducing the average
impedance $\langle
Z\rangle$ 
and 
variations 
$\delta Z (x{,}y){=} Z(x{,}y){-}\langle Z\rangle$, 
one should assume that the relative variations $\delta Z(x{,}y){/}
\langle Z \rangle $
 are of the 
 order of
$\xi$. In that case, the gauge
transformation (\ref{TzEqs-dn})-(\ref{InvTzEqs-dn}) allows 
spatial 
derivatives of $\delta Z (x,y)$ 
to be taken into account by the model:
\begin{equation}
\left|\frac {  \delta Z(x,y) }{\langle Z \rangle } \right |\alt
\xi, \quad \left | \frac {  \nabla \delta Z(x,y) }{k_z \langle Z
\rangle } \right |  \alt \xi^2. \label{dZ_Valid_0}
\end{equation}
Otherwise (if $\delta Z (x,y)/\langle Z \rangle {\sim}\xi^2$), in conditions of the paraxial approximation, the effect 
of
periodic
variations $\delta Z(x,y)$ 
cancels out in 
the gauge transformation (\ref{TzEqs-dn})-(\ref{InvTzEqs-dn}),
yielding thus 
expressions for a 
uniform dielectric material
[Eqs.~(\ref{TzEqs}) and (\ref{InvTzEqs})].
The conditions 
in terms of 
refractive index variations 
$\delta n(x,y)$ 
follow from
Eq.~(\ref{dZ_Valid_0})
by substitution $n \rightarrow Z$ since $Z(x,y){=}\mu/n(x,y)$.

Following the analogy between photonic crystals 
and semiconductors, 
it is interesting to observe that
the
ratio between 
two expressions in (\ref{dZ_Valid_0}) reads
\begin{equation}
\left| \frac {   \nabla \delta Z }{k_z  \delta Z } \right| \alt
\xi \ll 1,\quad \Bigl(\text{or} \quad \left| \frac { \nabla
\delta n }{k_z  \delta n } \right|
\ll 1 \Bigr).
\label{dZ2_effMass_Valid}
\end{equation}
%
%
%
This 
corresponds 
to the condition\cite{Pekar46,Luttinger,BonchBruevich90} of smooth potential variations 
$ \left| \frac {\Lambda \nabla \delta U }{  \delta U } \right|
{\ll}1$
%
%
in the \emph{effective mass method}
widely used in solid-state physics.
%
%
%
In the 
case of paraxial light propagation considered here, the 
condition (\ref{dZ2_effMass_Valid}) verifies 
the use of effective mass $m_0{=} \hbar k_z/c\langle n^{-1}
\rangle $ in analysis 
of photonic bands  
in 
the lateral propagation 
direction.
By substituting the gauge transformation (\ref{TzEqs-dn})
 and applying the operators
$-i\hbar
e_{3\alpha\beta}Z^{-b}\hat{\mathcal{E}}^{-1}_{\beta\gamma}Z^{b}$
and
$i\hbar Z^{a}\hat{\mathcal{E}}^{-1}_{\beta\alpha}Z^{-a}$ 
in the first and second equations (\ref{SysEqs1_nZ}),
both equations are 
converted into
a similar form 
with respect to
 spinor functions
$\hat{\mathbf{z}}{\times}\bm{\psi_{q\mathbf{k}}}$ and
$\bm{\psi_{q\mathbf{k}}}$: 
%
%
%
\begin{equation}
\begin{split}
&\frac{c\hbar k_z }{n(x,y)}
\hat{\mathbf{z}}{\times}\bm{\psi}_{q\mathbf{k}}%
-\langle n^{-1} \rangle\frac{c\hbar
\Delta_\perp }{2k_z}\hat{\mathbf{z}}{\times}\bm{\psi}_{q\mathbf{k}}\\
&\quad+ic\hbar\Bigl(b{+}1{-}\frac{1}{2\mu}\Bigr)[\bm{\nabla}(
n^{-1}){\times}\bm{\psi}_{q\mathbf{k}}]{=}\hbar\omega
_{q\mathbf{k}}\hat{\mathbf{z}}{\times}\bm{\psi}_{q\mathbf{k}},
\\
 &\frac{c\hbar k_z }{n(x,y)}
\bm{\psi}_{q\mathbf{k}}-\langle n^{-1}\rangle\frac{c\hbar
\Delta_\perp}{2k_z}\bm{\psi}_{q\mathbf{k}}\\
&\quad+i\hat{\mathbf{z}}c\hbar\Bigl(a{-}1{-}\frac{1}{2\mu}\Bigr)\left(\bm{\nabla}(
n^{-1})\cdot\bm{\psi}_{q\mathbf{k}}\right){=}\hbar\omega
_{q\mathbf{k}}\bm{\psi}_{q\mathbf{k}}.
\end{split}
\label{SysEqs1_H_wgd}
\end{equation}%
The operators applied to 
$\hat{\mathbf{z}}{\times}\bm{\psi_{q\mathbf{k}}}$ and
$\bm{\psi_{q\mathbf{k}}}$ (in the left-hand side) show different
longitudinal 
components (third term in each equation).
At the same time, 
the 
gauge transformation (\ref{TzEqs-dn}) 
implies transversality of spinor functions  ($\hat{\mathbf{z}} 
\bm{\psi}_{q\mathbf{k}}{=}0$), such that these longitudinal
terms 
cancel out
by adjusting the gauge transformation parameters $a$ and $b$ :
\begin{equation}
a=(1+2\mu)/2\mu, \quad b=(1-2\mu)/2\mu, \label{powers_ab}
\end{equation}
where $\mu$ is the magnetic permeability ($a{=} \frac 32$ and $b{=}{-}\frac 12$ 
in the
optical
spectrum range). 

The vector cross product $\hat{\mathbf{z}} {\times}
\bm{\psi}_{q\mathbf{k}}$ in the first
equation (\ref{SysEqs1_H_wgd})
is 
the result of spin operator\cite{Boiko07}
$\hat{s}_z{=}i\hat{\mathbf{z}}{\times}$ applied to the spinor $\bm{\psi}_{q\mathbf{k}}$. 
In the paraxial gauge transformation considered here %
(with $\hat{\mathbf{z}}\bm{\psi}{=}0$), the spinor functions
are invariant under the operator $\hat {s}_z^2$.
%
%
Therefore,
substituting 
parameters 
(\ref{powers_ab})
and taking a 
cross product of $\mathbf{\hat{z}}$ and
first equation,
one obtains in 
(\ref{SysEqs1_H_wgd}) 
two
identical eigenproblems 
that read 
%
%
%
%
\begin{equation}
\begin{split}
&\!\left[ m_{0}c^2 \Bigl\langle \frac 1 n \Bigr\rangle^2+\frac{
\mathbf{\hat{p}}_{\bot }^{2}}{2m_{0}} \right.\\
&\quad\quad{+} \left. m_{0}c^2 \Bigl\langle \frac 1 n
\Bigr\rangle \Bigl( \frac 1{n(x,y)} {-} \Bigl\langle \frac1n
\Bigr\rangle \Bigr)
%
%
%
\right] \bm{\psi}_{q\mathbf{k}}{=}\hbar \omega _{q\mathbf{k}}
\bm{\psi}_{q\mathbf{k}}
%
%
\end{split}
\label{H_wgd}
\end{equation}%
where $m_0{=} \hbar k_z/c\langle n^{-1} \rangle $
is the effective mass and $\mathbf{\hat{p}}_{\bot }=-i\hbar
\mathbf{\nabla }_{\bot }$ is the momentum operator in the lateral
($xy$-plane) direction. The third term in the Hamiltonian
(\ref{H_wgd}) is the effective crystal potential induced by
variations of the (complex) refractive index $n(x,y)$.

This is the stationary Schr\"odinger equation for
photons in dissipative dielectric lattices of coupled
microcavities or parallel waveguides.
%
The orthogonality relationship
%
can be established
between the concomitant partners 
of biorthonormal set of its solutions 
[see
Eq.~(\ref{VmkNorm})]
%
%
\begin{eqnarray}
\hspace{-0.1in}(\tilde{\psi}_{q^{\prime }\mathbf{k}^{\prime
}}^{(\alpha)}| \psi_{q\mathbf{k}}^{(\alpha)} ) {=} \int
(\bm{\tilde{\psi}}_{q^{\prime } \mathbf{k}^{\prime }}^{*}{\cdot}
\bm{\psi}_{q\mathbf{k}}) d^{2}\mathbf{r}_{\bot } {=}\delta \left(
\mathbf{k} _{\bot }-\mathbf{k}_{\bot }^{\prime }\right) \delta
_{q^{\prime }q}. \label{VmkNorm_dn}
\end{eqnarray}%

In the 
particular case of parallel dielectric waveguides, the
conventional waveguiding theories
have established 
the mode orthogonality relationship in terms of the electric and
magnetic field components.
The biorthonormal set of solutions 
in mirror-patterned structures
is shown to satisfy 
such relationship [Sec.~\ref{SecMirrorPatterned},
Eq.~(\ref{ORT1})].
However, the paraxial gauge transformation in 
periodic
dielectric structures [Eq.(\ref{TzEqs-dn})] differs from the one
in mirror-patterned microcavities with uniform dielectric
material in the cavity [Eq.(\ref{TzEqs})]. Therefore, it is
crucial to
verify that the 
gauge transformation (\ref{TzEqs-dn}) and the biorthonormal
orthogonality relationship (\ref{VmkNorm_dn})
%
are in agreement with the well-established results of
conventional waveguiding theories.

%
%
%

An 
electromagnetic wave 
associated with the concomitant partner
$\tilde{\bm{\psi}}_{q\mathbf{k}}$ [in (\ref{VmkNorm_dn})]
propagates in the same structure as the wave 
(\ref{TzEqs-dn}) 
but with the complex conjugated refractive index and impedance:
%
%
%
%
\begin{equation}
\begin{split}
\tilde{\mathbf{e}}_{q\mathbf{k}}&{=} (Z^{a})^{*}
\hat{\bm{\mathcal{E}}}
(Z^{-a} )^{*} {\cdot} \tilde{\bm{\psi}}_{q\mathbf{k}}, \quad \\
\tilde{\mathbf{h }}_{q\mathbf{k}}&{=} (Z^{-b})^{*}
\hat{\bm{\mathcal{E}}} (Z^{b})^{*} {\cdot}
  \mathbf{\hat{z}}{\times}
\tilde{\bm{\psi}}_{q\mathbf{k}}.
\end{split}
\end{equation}
Respectively, the inverse transformation reads
\begin{equation}
\begin{split}
\tilde{\bm{\psi}}_{q\mathbf{k}}&{=} (Z^{a})^{*}
\hat{\bm{\mathcal{E}}}^{-1}
(Z^{-a})^{*} {\cdot}  \tilde{\mathbf{e}}_{q\mathbf{k}}, \quad \\
\mathbf{\hat{z}}{\times} \tilde{\bm{\psi}}_{q\mathbf{k}} &{=}
(Z^{-b})^{*}
\hat{\bm{\mathcal{E}}}^{-1} (Z^{b})^{*} {\cdot} \tilde{\mathbf{h
}}_{q\mathbf{k}}.
\end{split}
 \label{inv_concom_gauge_dn}
\end{equation}
Substituting $-\mathbf{\hat{z}}{\times}[\mathbf{\hat{z}}{\times}
\tilde{\bm{\psi}}_{q^\prime\mathbf{k}^\prime} ]$
[Eq.(\ref{inv_concom_gauge_dn})] and $\bm{\psi}_{q\mathbf{k}}$
[Eq.(\ref{InvTzEqs-dn})] in the orthogonality relationship
(\ref{VmkNorm_dn}),
we obtain
%
\begin{equation}
\begin{split}
&\delta \left( \mathbf{k} _{\bot }-\mathbf{k}_{\bot }^{\prime
}\right) \delta _{q^{\prime }q}
%
%
{=}{-}\!\!\!\int\!\!\hat{\mathbf{z}}{\cdot}
\!\Bigl(\!\hat{\bm{\mathcal{E}}}^{-1}\tilde{\mathbf{h
}}_{q^\prime\mathbf{k}^\prime}\!\Bigr)^{\!*} \!\! {\times}\!
\Bigl(\!\hat{\bm{\mathcal{E}}}^{-1}\!\!\mathbf{e
}_{q\mathbf{k}}\!\Bigr) d^2 \mathbf{r}_\perp
 \\
&\quad \quad \quad  {+}ib \int
\hat{\mathbf{z}}{\cdot}[\hat{\mathbf{z}}{\times}(\hat{\bm{\mathcal{E}}}^{-1}\mathbf{e
}_{q\mathbf{k}})] \frac{(\nabla(Z){\cdot}\tilde{\mathbf{h
}}^{*}_{q^\prime\mathbf{k}^\prime})}{k_zZ} d^2 \mathbf{r}_\perp\\
&\quad \quad \quad {-} ia \int \hat{\mathbf{z}}{\cdot}
[(\hat{\bm{\mathcal{E}}}^{-1}\tilde{\mathbf{h
}}_{q^\prime\mathbf{k}^\prime})^{*} {\times} \hat{\mathbf{z}}
 ] \frac{(\nabla(Z){\cdot}\mathbf{e
}_{q\mathbf{k}})}{k_zZ} d^2 \mathbf{r}_\perp{.}
\end{split}
\label{INterimGG}
\end{equation}
The second and third terms in the right-hand side of this
equation are null.
The first term 
coincides with expression in Eq.~(\ref{NormhEepsilon_Inv}), 
yielding
the orthogonality relationship 
[see Eq.~(\ref {ORT1_eh})] 
\begin{equation}
\int \mathbf{\hat{z}} {\cdot} [\mathbf{\tilde{h}}_{q^{\prime
}\mathbf{k}^{\prime }}^{*} {\times} \mathbf{e }_{q\mathbf{k}}]
d^{2}\mathbf{r}_{\bot } {=} - \delta \left( \mathbf{k}_{\bot }-
\mathbf{k}_{\bot }^{\prime
}\right) \delta _{q^{\prime }q}.  
\label{ORT1_eh_dn}
\end{equation}
The fields $\mathbf{\tilde{E}}_{q^{\prime }\mathbf{k}^{\prime }}$
and
$\mathbf{\tilde{H}}_{q^{\prime }\mathbf{k}^{\prime }}$ 
of the wave associated with the concomitant partner
$\tilde{\bm{\psi}}_{q^\prime\mathbf{k}^\prime}$ are defined by
relationship (\ref{SolutionForm_dn})
with the 
complex conjugated dielectric function. 
Taking this fact into account, one can 
verify that
\begin{equation}
\int \mathbf{\hat{z}} {\cdot} [\mathbf{\tilde{H}}_{q^{\prime
}\mathbf{k}^{\prime }}^{*} {\times} \mathbf{E}_{q\mathbf{k}}]
d^{3}\mathbf{r}_{\bot } {=} {-} 
\delta (k_{z}{-} k_{z}^{\prime })\delta \left( \mathbf{k}_{\bot
}{-} \mathbf{k}_{\bot }^{\prime }\right) \delta _{q^{\prime }q},
\label{ORT1_dn}
\end{equation}
where 
integration runs over the entire structure
of parallel dielectric waveguides (or 
cavity-unfolded array of 
microcavities.)
%
%
The 
expression (\ref{ORT1_dn}) corresponds to 
conventional 
orthogonality relationship between the modes of parallel
dielectric
waveguides.
This result verifies 
the theoretical treatment presented in this paper.



For 
low-order photonic bands,
the 
structure parameters
satisfying
the 
paraxial approximation conditions
can be defined 
more precisely. 
Thus, for a particular photonic band, the conditions
(\ref{dZ_Valid_0})
expressed in terms of intraband matrix elements read
\begin{equation}
\frac { | \langle \tilde{\psi}_{q\mathbf{k}}^{(\alpha)}| \delta n|
{\psi}_{q\mathbf{k}}^{(\alpha)} \rangle|}{|\langle n \rangle |}
\alt \xi, \quad \frac { | \langle
\tilde{\psi}_{q\mathbf{k}}^{(\alpha)}| \nabla \delta n |
{\psi}_{q\mathbf{k}}^{(\alpha)} \rangle|}{k_z |\langle n \rangle
|} \alt \xi^2, \label{dn_MatrEq_Valid}
\end{equation}
where the relationship $Z(x,y){=}\mu/n(x,y)$ is taken into account. 
Being applied to the low-order 
bands, 
these 
expressions yield 
the relationships (\ref{dn_valid}) [see also the discussion in
Sec.~\ref{SubSecHamiltonian}].

%
%

%
%
%
%
%
%
%


\section{Results and discussion} 

\label{PropSol}
\label{ResultsAndDiscussion}

\subsection{Generalized Hamiltonian} 
\label{GenHamiltonian}


Combining 
the results of
Sec.~\ref{SecMirrorPatterned}
and 
\ref{SecIndexPatterned} [Eqs.~(\ref{H}) and (\ref{H_wgd})],
we obtain 
the Hamiltonian for 
entire subclass of 2D 
structures 
characterized by paraxial
light propagation 
in the direction 
normal to %
periodic 
lattice plane
\begin{equation}
\hspace{-0.05in}\begin{split}%
&\left[ m_{0}c^2 \Bigl\langle \frac 1 n \Bigr\rangle^2+\frac{
\mathbf{\hat{p}}_{\bot }^{2}}{2m_{0}} +i\Bigl\langle \frac 1n \Bigr\rangle \frac{c\hbar }{2L}\ln R(x,y) \right.\\%
&\quad\quad{+} \left. m_{0}c^2 \Bigl\langle \frac 1 n
\Bigr\rangle \Bigl( \frac 1{n(x,y)} {-} \Bigl\langle \frac1n
\Bigr\rangle \Bigr)
%
%
%
\right] \bm{\psi}_{q\mathbf{k}}{=}\hbar \omega _{q\mathbf{k}}
\bm{\psi}_{q\mathbf{k}}. \label{H_all}
%
%
\end{split}
\end{equation}%
The photonic state wave function $\bm{\psi}_{q\mathbf{k}}$
(spinor) is
related 
with 
the corresponding electromagnetic wave
via paraxial gauge transformation
\begin{equation}
\hspace{-0.05in}\left[
\begin{matrix}
E_{q\mathbf{k}}^{(\alpha)} \\
H_{q\mathbf{k}}^{(\gamma)}
\end{matrix}
  \right]
 \hspace{-0.05in}
  {=}e^{ik_z z{-}i\omega t} \frac{1{+}\eta(z)}{\sqrt{2\pi}}
\hspace{-0.04in} \left[
\begin{matrix}
Z^{\frac12+a} \hat{\mathcal{E}}_{\alpha\beta}Z^{-a} \\
Z^{-\frac12-b}e_{3\beta\alpha}\hat{\mathcal{E}}_{\gamma\alpha}Z^{b}
\end{matrix}
\right]\hspace{-0.04in} \psi_{q\mathbf{k}}^{(\beta)}, 
\label{Gauge_all}
\end{equation}
where operator $\bm{\hat{\mathcal{E}}}$ was 
introduced in Eq.~(\ref{OperatorE}) and parameters $a$ and $b$
were obtained in Eq.~(\ref{powers_ab}).
%


The effective crystal potential 
is defined 
by the third and fourth terms in the left-hand side of Eq.~(\ref{H_all}).  
%
In the case of 2D arrays of coupled microcavities, it
takes into account the 
effects of mirror reflectivity patterning, as
in metal-patterned VCSEL arrays\cite{Orenstein91}
(Fig.~\ref{fig1}),
and dielectric material 
variations, as 
in periodically etched 
VCSEL structures\cite{Gourley91,Mawst03} (Fig.~\ref{fig2}).
%
Such simple expression for the effective potential
was obtained 
by 
unfolding 
the cavities 
along the optical 
axis [$z$-axis in Fig.~\ref{fig1}(d)] and representing 
the standing optical modes 
in the form of propagating
Bloch waves (\ref{Gauge_all}) 
in equivalent 3D structure. 

In a periodic array of microcavities, the longitudinal component
of wave
vector ($k_z$) 
is fixed by the self-repetition condition at the cavity
roundtrip. At the same time, Eq.~(\ref{H_all})
cannot reproduce the 
cavity
resonance 
condition.
  Since
Eq.~(\ref{H_all}) takes into account a phase shift at the cavity
mirrors, 
$k_z$
has to
be evaluated from 
the roundtrip
condition in a cavity with perfectly reflecting mirrors. In
particular,
$k_z{=}2 \pi /L$ in the case of one-wavelength microcavities 
($L{=}\lambda/n$).
%
%
%
%
In addition, analyzing a 
symmetry of
the group of 
$\mathbf{k}$,
the $z$-axis nonreciprocity 
of equivalent 
cavity-unfolded structure 
has to be
taken into account 
(see Sec. \ref{SecMirrorPatterned}).
%
%
%
%
Obviously, for an 
array of parallel dielectric waveguides or antiguides (\emph{e.g.}, 
photonic band gap fibers,\cite{Argyros05} 
photonic liquid crystal fibers\cite{Wolinski06}),
these restrictions 
of the model 
do not apply.
%
The effective potential
%
is uniquely defined 
by the refractive index 
variations [fourth term in the Hamiltonian of Eq.(\ref{H_all})].

%
%
For most important (in practical applications) low-order bands,
and within the range of
parameters 
limited by conditions (\ref{dR_valid}) and (\ref{dn_valid}), the
Hamiltonian 
is suitable for 
structures with high contrast of refractive index variations
(\emph{e.g.}, 
etched
arrays of 
pillar microcavities,\cite{Bayer99} 
holey photonic crystal fibers\cite{Russell03}).
The class of 2D photonic crystal materials encompassing 
valid
solutions of the non-Hermitian Hamiltonian eigenproblem
(\ref{H_all}) is further illustrated below with several 
structure examples.

Arrays 
of semiconductor microcavities 
($n{=}3.5$) defined by mirror reflectivity patterning
(Fig.\ref{fig1}) and operating at the optical wavelength 
$\lambda{\sim} 1 $ $\mu m$ typically
employ 
lattices of the pitch $\Lambda{\sim} 5~\mu m$. 
%
%
These parameters assume that 
$\xi^2{\sim} 10^{-3}$ [Eq.~(\ref{dR_valid})] and for any lattice
fill factor $FF$,
the 
Hamiltonian 
can be applied
to 
the structures with mirror reflectivity 
contrast $\delta R $ up to
$10^{-2}$.
Note that
in a typical VCSEL array with mirror reflectivity patterning,
the reflectivity contrast 
is in the range from ${\sim} 10^{-4}$ to ${\sim} 10^{-3}$.





In periodic dielectric structures 
exhibiting
high-contrast variations
$|\delta n| {\agt}
 1$ (arrays of micropillars\cite{Bayer99} or holey photonic
crystal fibers\cite{Russell03}),
the paraxial Hamiltonian 
(\ref{H_all}) 
%
applies in the two opposite cases of lattice cell parameters
[see Eq.~(\ref{dn_valid})], at $FF \lesssim \lambda /2 \Lambda |\delta n|$ 
(low 
fill factor) or at $FF \gtrsim 1{-} \lambda /2 \Lambda |\delta n
|$
(high
fill factor).
%
%
Thus, for 
a $5~\mu m$-pitch array 
of deeply etched semiconductor microcavity pillars ($n{=}3.5$,
$|\delta n|{=}2.5$)
%
%
operating at 
${\sim} 1~\mu m$ wavelength,
%
%
low-order photonic bands 
can be treated by the model 
in the case of lattices with fill factor $FF{<}0.2$ or $FF{>}0.8$.
%
%



%

The first working silica photonic crystal fiber\cite{Russell03}
has a cladding material consisting of $300$ $nm$ air holes
arranged
in a hexagonal lattice of $2.5$ $\mu m$ pitch. 
The refractive index of fused silica varies from $1.55$ to $1.44$
in the 
wavelength range of $0.2 - 1.5~\mu m$. The lattice fill factor
and 
the contrast  are thus $FF{\sim}0.02$ and $|\delta n|{\sim}0.5 $.
%
%
Due to the small 
size of air holes,
 the 
average 
refractive index of the structure 
is close to that 
of 
fused silica [$\langle 1/n \rangle^{-1} {=}(1{+}|\delta
n|)/(1{+}FF|\delta n|) {\sim} 1.5 $ in Eq.~(\ref{dn_valid})].
For such 
holey photonic crystal fibers, 
the 
Hamiltonian (\ref{H_all}) is
accurate throughout the entire optical transparency 
range of fused silica, 
from ultraviolet ($\lambda{\sim}200~nm$) to infrared
($\lambda{\sim}1.5~\mu m$) wavelengths.
%
%
%
%

In the case of 
arrays 
of parallel dielectric waveguides (or antiguides) 
with
low-contrast refractive index variations 
($|\delta n| {\ll} 1$),
the 
Hamiltonian
(\ref{H_all}) is valid for 
any fill factor of the lattice,
in the optical wavelength range 
from $\lambda{\sim} 4|\delta n| \Lambda $ [Eq.(\ref{dn_valid})]
to $ \lambda{\sim} 0.2 n \Lambda  $.
(The long wavelength range is limited by 
the paraxial approximation condition $\xi {\ll} 1 $ 
with 
the critical value
of $\xi {\sim} 0.1 $ .)
The 
silica photonic band gap fibers reported in
Ref.~\onlinecite{Argyros05} ($n{\sim} 1.5$, $\delta n {=}0.015$
and $\Lambda{=}6~ \mu m$) satisfy 
the paraxial approximation
conditions 
in the visible ($\lambda{>}360~nm$) and near infrared regions of
the optical spectrum,
up to 
the upper transparency edge of fused silica ($\lambda{\sim}1.5~\mu
m$).

\subsection{Biorthonormal 
solutions 
in 
lattices with inversion symmetry}

The Hamiltonian in Eq.(\ref{H_all}) is independent of the spin
variables. 
Therefore, all states are 
two-fold degenerate by spin (polarization), yielding thus
degeneracy of the biorthonormal spinors
$\bm{\psi}_{q\mathbf{k}\uparrow}(t,\mathbf{r}_\perp)$ and
$\bm{\psi}_{q\mathbf{k}\downarrow}(t,\mathbf{r}_\perp)$
associated with the eigenvalues $\hbar \omega
_{q\mathbf{k}\uparrow}=\hbar \omega_{q\mathbf{k}\downarrow}$.
(Arrows indicate the spin direction.)\cite{Kittel}
This fact allows the 
Eq. (\ref{H_all})
to be transformed into a scalar eigen problem with respect to the
amplitudes of positive-spin (negative-spin) components of the
spinors
$\bm{\psi}_{q\mathbf{k}{\uparrow}{,}{\downarrow}}(t{,}\mathbf{r}_\perp){=}\psi_{q\mathbf{k}}(t{,}\mathbf{r}_\perp
)|{\uparrow}{,}{\downarrow} \rangle $ and 
$\bm{\tilde{\psi}}_{q\mathbf{k}{\uparrow}{,}{\downarrow}}(t{,}\mathbf{r}_\perp)=\tilde{\psi}_{q\mathbf{k}}(t{,}\mathbf{r}_\perp
)|{\uparrow},{\downarrow} \rangle $  .


In the case of 2D photonic lattices 
exhibiting inversion symmetry 
(\emph{e.g.}, square or triangular lattices), the  biorthonormal
system of lattice-periodic 
functions
\begin{eqnarray}
\mathbf{\psi}_{q\mathbf{k}\uparrow,\downarrow}
(t,\mathbf{r}_\perp) &=& e^{-i\omega
t+i\mathbf{k}_\perp\mathbf{r}_\perp}\mathbf{u}_{q\mathbf{k}\uparrow,\downarrow}
(\mathbf{r}_\perp), \label{ketBlochSpinor}
\\
\mathbf{\tilde{\psi}}_{q\mathbf{k}\uparrow,\downarrow}
(t,\mathbf{r}_\perp) &=& e^{-i\omega^{*}
t+i\mathbf{k}_\perp\mathbf{r}_\perp}\mathbf{\tilde{u}}_{q\mathbf{k}\uparrow,\downarrow}
(\mathbf{r}_\perp) \label{braBlochSpinor}
\end{eqnarray}
can be readily obtained by applying the 
$PT$ transformation (time reversal followed by coordinate
inversion) to
Eq.(\ref{H_all}) 
and  noting 
the degeneracy $\hbar \omega
_{q\mathbf{k}\uparrow}=\hbar \omega_{q\mathbf{k}\downarrow}$.
This has much in common with the Kramers
degeneracy\cite{Kramers30,Wigner32} of a single-electron
Hamiltonian in 
lattices with inversion symmetry.
The invariance of the
single-electron 
Hamiltonian
 under the $PT$
transformation 
yields
degeneracy of eigenvalues
$E_{q\mathbf{k}\uparrow}=E_{q\mathbf{k}\downarrow}$
associated with the states of opposite spin. 
%
%


In the Hamiltonian (\ref{H_all}), the degeneracy on spin variable 
($\hbar \omega _{q\mathbf{k}\uparrow}{=}\hbar
\omega_{q\mathbf{k}\downarrow}$) is caused by the fact that 
the Hamiltonian contains 
no spin operators. 
On the other hand, the Hamiltonian itself 
is
not invariant under the $PT$ transformation. In the lattices with
inversion symmetry, its $PT$ transform
is uniquely defined 
by the result of the time reversal operation (complex conjugation 
followed by substitution ${-} t \rightarrow
t$),\cite{LandauIII,LandauIV} which
transforms $\hat H$ into $\hat H^{*}{=}\hat H^+$.
%
%

The $PT$ transformation of the spinor functions
(\ref{ketBlochSpinor})
reads
\begin{equation}
\begin{split}
& PT :  \bm{\psi}_{q\mathbf{k}\uparrow,\downarrow}
(t,\mathbf{r}_\perp)
 = e^{-i\omega
t+i\mathbf{k}_\perp\mathbf{r}_\perp}\mathbf{u}_{q\mathbf{k}}
(\mathbf{r}_\perp)
\\
&{\rightarrow}
{-}\bm{\psi}^{*}_{q\mathbf{k}{\downarrow}{,}{\uparrow}}
({-}t{,}{-}\mathbf{r}_\perp)  {=} {-} e^{{-}i\omega^{*}
t{+}i\mathbf{k}_\perp\mathbf{r}_\perp}\mathbf{u}^{*}_{q\mathbf{k}{\downarrow}{,}{\uparrow}}
({-}\mathbf{r}_\perp)
\end{split}
\label{PTketBlochSpinor}
\end{equation}
where 
transformation alters the spin
direction\cite{Kramers30,Wigner32,InuiTanabe}
 (for $s{=}1$,
$|{\uparrow},\downarrow\rangle^{*}{=}{-}|{\downarrow},\uparrow
\rangle$ in accordance with the phase convention of
Refs.~\onlinecite{LandauIII,LandauIV}).
Therefore, the $PT$ transform of 
Eq.(\ref{H_all})
with 
the Hamiltonian exhibiting the 
symmetry $\hat H^{*}{=}\hat H^+$ and $\hat H (-\mathbf{r}_\perp)
{=}\hat H (\mathbf{r}_\perp)$
yields, within the accuracy of a phase factor, 
\begin{equation}
\begin{split}
PT  \: : \:  \: \hat H \bm{\psi}_{q\mathbf{k}\uparrow,\downarrow}
(t,\mathbf{r}_\perp)&{=}\hbar
\omega_{q\mathbf{k}\uparrow,\downarrow} \bm{\psi}_{q\mathbf{k}\uparrow,\downarrow} (t,\mathbf{r}_\perp) \\
 \rightarrow \:
 \: \hat H^{+} \bm{\psi}^{*}_{q\mathbf{k}\downarrow,\uparrow}
({-}t,{-}\mathbf{r}_\perp)&{=}\hbar
\omega^{*}_{q\mathbf{k}\uparrow,\downarrow}
\bm{\psi}^{*}_{q\mathbf{k}\downarrow,\uparrow}
({-}t,{-}\mathbf{r}_\perp){,}
\end{split}
\label{PTscalarHPsi}
\end{equation}
%
where $\hbar\omega^{*}_{q\mathbf{k}\uparrow}{=}\hbar
\omega^{*}_{q\mathbf{k}\downarrow}$. Comparison of
with 
Eqs.~(\ref{eigH})-(\ref{eigHdag})
shows
that 
the co-partners of biorthonormal set
can be chosen as
\begin{equation}
\begin{split}
\bm{\psi}_{q\mathbf{k}{\uparrow}{,}{\downarrow}}
(t{,}\mathbf{r}_\perp)
&{=} \psi_{q\mathbf{k}}(t,\mathbf{r}_\perp) |{\uparrow}{,}{\downarrow}\rangle \\
\bm{\tilde \psi}_{q\mathbf{k}{\uparrow}{,}{\downarrow}}
(t{,}\mathbf{r}_\perp)
&{=}\bm{\psi}^{*}_{q\mathbf{k}{\uparrow}{,}{\downarrow}}
({-}t{,}{-}\mathbf{r}_\perp){=}\psi^{*}_{q\mathbf{k}}({-}t{,}{-}\mathbf{r}_\perp)
|{\uparrow}{,}{\downarrow}\rangle. \label{Bra&Kets}
\end{split}
\end{equation}
where $\psi_{q\mathbf{k}}(t,\mathbf{r}_\perp)$ and its $PT$
transform $\tilde{\psi}_{q\mathbf{k}}(t,\mathbf{r}_\perp)=\psi^{*}_{q\mathbf{k}}({-}t{,}{-}\mathbf{r}_\perp)$ are 
the 
scalar amplitudes of 
nonzero spinor component and its co-partner. 
In periodic lattices, these are 
the periodic Bloch waves
with 
plane wave 
envelopes 
related 
via the $PT$-transform [see 
Eqs.~(\ref{ketBlochSpinor})-(\ref{braBlochSpinor})].

In 
(\ref{Bra&Kets}), the relative phase 
of 
co-partners $\bm{\psi}_{q\mathbf{k}\uparrow,\downarrow}$ and
$\bm{\tilde{\psi}}_{q\mathbf{k}\uparrow,\downarrow}$ 
is set 
by the 
orthogonality
relationship (\ref{VmkNorm2}). This relationship 
evidences  
that 
an arbitrary phase factor $e^{i\alpha}$ can also be 
introduced in both equations (\ref{Bra&Kets}),  modifying thus
the absolute phases of 
co-partners but preserving their 
relative phase 
shift. 
The fact that the co-partner phases have to be 
the same (at $t{=}0$) 
can be seen 
from 
the following observation as well: 
There should be no difference
between the co-
partners 
in the case of Hermitian Hamiltonian. Therefore, the functions
$\bm{\psi}_{q\mathbf{k}\uparrow,\downarrow}(t,\mathbf{r}_\perp)$
and
$\bm{\tilde{\psi}}_{q\mathbf{k}\uparrow,\downarrow}(t,\mathbf{r}_\perp)$
coincide
in the limit $ \text{Im} (\omega_{q\mathbf{k}}) {\rightarrow} 0 $.

%
%
%
%

Eq.~(\ref{Bra&Kets}) defines the biorthonormal system
of solutions
in the case of lattices with inversion symmetry. It
effectively transforms (\ref{H_all}) into a scalar Hamiltonian
eigenproblem
with respect to 
the amplitude $\psi_{q\mathbf{k}}
(t,\mathbf{r}_\perp)$ of nonzero spinor component. 
The orthogonality relationship for the biorthonormal set of
scalar functions 
$\psi_{q\mathbf{k}}(t,\mathbf{r}_\perp)$ and
$\tilde{\psi}_{q\mathbf{k}}(t{,}\mathbf{r}_\perp){=}\psi^{*}_{q\mathbf{k}}({-}t{,}{-}\mathbf{r}_\perp)$
follows 
from Eq.~(\ref{VmkNorm2}): 
\begin{equation}
\begin{split}
&\langle \tilde{\psi}_{q^{\prime}\mathbf{k}}| \psi_{q\mathbf{k}}
\rangle \\
&~~~~~{=} \frac{\left( 2\pi \right) ^{2}}{ \Omega _{\bot
}}{\int_{\text{cell}}}
\psi_{q^{\prime }\mathbf{k}}({-}t{,}{-}\mathbf{r}_\perp)
\psi_{q\mathbf{k}}(t{,}\mathbf{r}_\perp) d^{2}\mathbf{r}_{\bot }
{=} \delta _{q^{\prime}q}.
\end{split}
\label{VmkNorm2_ScalarInvSym}
\end{equation}%

Finally, note 
that in the case of significant difference between reflection
coefficients of a mirror for $s$- and $p$-polarized waves
in 
microcavities
( $|\ln R_s/R_p| {\gtrsim} \xi^2$ in (\ref{reflection})) 
or large 
polarization anisotropy in periodic 
dielectric lattices
($|n_x{-}n_y|\langle n^{-1}\rangle {\gtrsim} \xi^2$), 
%
the spin-orbit coupling effects set in, 
rendering invalid
the scalar approximation
(\ref{Bra&Kets}). 
Another example of 
spin degeneracy removal,
for which 
the scalar approximation (\ref{Bra&Kets}) 
is
unsuitable, is the Coriolis-Zeeman splitting of photonic energy bands 
in
nonpermanent gravitational field.\cite{Boiko07}



\subsection{Biorthonormal plane wave expansion 
in square 
lattices}

For square lattice structures 
(Sec.~\ref{NumRes}), it is more convenient to express
the biorthonormal set of spinor functions (\ref{Bra&Kets})
in the Cartesian coordinates basis,
in the form of functions
$\bm{\psi}{=}{\Bigl(}\begin{smallmatrix} \psi^{(x)} \\
\psi^{(y)} \end{smallmatrix} {\Bigr)}$. 
For a first rank tensor $\bm{\psi}$ representing a state of spin
$s{=}1$, the relationships\cite{LandauII}
$\psi^{(x)}{=}i(\psi_{1,1}{-}\psi_{1,-1}){/}\sqrt{2}$ and
$\psi^{(y)}{=}(\psi_{1,1}{+}\psi_{1,-1}){/}\sqrt{2}$ provide 
a transformation between 
the Cartesian coordinates 
and
$|s,m_s\rangle$ 
functions 
bases.
(For paraxial
photonic states considered here,
the spinor component $\psi^{(z)}{=}{-}i\psi_{1{,}0}$ 
is null and therefore not indicated explicitly.)

The degenerate 
spinor functions (\ref{Bra&Kets}) 
can be
represented 
as linear combinations of
positive- and negative-spin states 
\begin{equation}
\begin{split}
\bm{\psi}_{q\mathbf{k}{,}\mathbf{\hat x}}&{=}
\frac{{-}i{(}\bm{\psi}_{q\mathbf{k}{\uparrow}}{-}\bm{\psi}_{q\mathbf{k}{\downarrow}}{)}}
{\sqrt{2}}
{,}~~
\bm{\tilde{\psi}}_{q\mathbf{k}{,}\mathbf{\hat
x}}{=}\frac{{-}i{(}\bm{\tilde{\psi}}_{q\mathbf{k}{\uparrow}}{-}\bm{\tilde{\psi}}_{q\mathbf{k}{\downarrow}}{)}}
{\sqrt{2}}
{,}
\\
\bm{\psi}_{q\mathbf{k},\mathbf{\hat y}}&{=}\frac
{\bm{\psi}_{q\mathbf{k}\uparrow}{+}\bm{\psi}_{q\mathbf{k}\downarrow}}
{\sqrt{2}}
{,}~~~~~~~~
\bm{\tilde{\psi}}_{q\mathbf{k},\mathbf{\hat y}}{=}\frac
{\bm{\tilde{\psi}}_{q\mathbf{k}\uparrow}+\bm{\tilde{\psi}}_{q\mathbf{k}\downarrow}}
{\sqrt{2}}
,
\end{split}
\end{equation}
yielding
the biorthonormal set of solutions
associated with the $x$- and $y$-polarized states of
electromagnetic field 
%
%
\begin{equation}
\begin{split}
\bm{\psi}_{q\mathbf{k}{,}\mathbf{\hat{x}}} 
&{=}
\biggl(
\begin{matrix} \psi_{q\mathbf{k}} (t{,}\mathbf{r}_\perp)\\
0
\end{matrix}
\biggr)
{,}~~~
\bm{\tilde{\psi}}_{q\mathbf{k}{,}\mathbf{\hat{x}}} 
{=}
\biggl(
\begin{matrix} \psi^{*}_{q\mathbf{k}} ({-}t{,}{-}\mathbf{r}_\perp)\\
0
\end{matrix}
\biggr)
{,}
\\
\bm{\psi}_{q\mathbf{k}{,}\mathbf{\hat{y}}}
&{=}
%
\biggl(
\begin{matrix}
0 \\
\psi_{q\mathbf{k}} (t{,}\mathbf{r}_\perp) \end{matrix} \biggr)
{,}~~~
\bm{\tilde{\psi}}_{q\mathbf{k}{,}\mathbf{\hat{y}}}
{=}
%
\biggl(
\begin{matrix}
0 \\
\psi^{*}_{q\mathbf{k}} ({-}t{,}{-}\mathbf{r}_\perp) \end{matrix}
\biggr){.}
\end{split}
\label{Bra&ketsX&Y}
\end{equation}
Here, the spinor components 
are expressed in the basis of Cartesian coordinates. Like 
the 
functions 
(\ref{Bra&Kets}), 
degenerate spinor functions 
$\bm{\psi}_{q\mathbf{k},\mathbf{\hat x}}$ and
$\bm{\psi}_{q\mathbf{k},\mathbf{\hat y}}$ 
($\hbar \omega_{q\mathbf{k},\mathbf{\hat{x}}}{=}\hbar
\omega_{q\mathbf{k},\mathbf{\hat{y}}}{=}\hbar
\omega_{q\mathbf{k}\uparrow,\downarrow}$) 
convert Eq.~(\ref{H_all}) into 
an eigenproblem with respect to 
scalar amplitudes
$\psi_{q\mathbf{k}}(t,\mathbf{r}_\perp)$, 
which assume the orthogonality relationship
(\ref{VmkNorm2_ScalarInvSym}).

For nonzero 
spinor components, the 
orthogonally polarized states 
(\ref{Bra&ketsX&Y})
%
%
show
equal 
distributions 
$|\psi_{q\mathbf{k}}(t,\mathbf{r}_\perp)|^2$ associated with the
 energy flux along the $z$ axis.
%
%
In experiment, such states 
are 
observed as orthogonally polarized modes 
showing 
indistinguishable 
intensity patterns.\cite{Boiko04,Guerrero04B} 
%
%
%
%
%

The 
%
stationary Schr\"odinger equation 
(\ref{H_all})
is solved 
here 
using a biorthonormal plane wave expansion of the periodic crystal
potential
and Bloch functions
(\ref{Bra&ketsX&Y}). In the 
stationary case,
the time
evolution 
of 
wave 
functions 
can be omitted, yielding the biorthonormal set of scalar
amplitudes 
$\psi_{q\mathbf{k}} (\mathbf{r}_\perp)$ and
$\tilde{\psi}_{q\mathbf{k}}
(\mathbf{r}_\perp){=}\psi^{*}_{q\mathbf{k}}
({-}\mathbf{r}_\perp)$.
The relationship between co-partners of the set 
implies
complex conjugated coefficients of 
expansion:
\begin{equation}
\begin{split}
\psi_{q\mathbf{k}}(\mathbf{r}_\perp)
& =\frac{1}{2\pi} 
e^{i\mathbf{k}_\perp\mathbf{r}_\perp}
 \sum_{\mathbf G}
C_{q\mathbf{k}}(\mathbf{G}) \exp( i\mathbf
{Gr}_\perp),\\
\tilde{\psi}_{q\mathbf{k}}(\mathbf{r}_\perp)
& =\frac{1}{2\pi}
e^{i\mathbf{k}_\perp\mathbf{r}_\perp}
 \sum_{\mathbf G}
C_{q\mathbf{k}}^{*}(\mathbf{G}) \exp( i\mathbf {Gr}_\perp),
\end{split}
\label{Psi_OPWs}
\end{equation}
where $\mathbf G$ is reciprocal lattice vector in the $x$-$y$
plane.
The
difference in 
expansion coefficients 
in non-Hermitian and Hermitian Hamiltonian cases 
can be appreciated 
by examining the Parseval
theorem   
for
coefficients $C_{q\mathbf{k}}(\mathbf{G})$:
%
\begin{equation}
\langle \tilde{\psi}_{q^{\prime }\mathbf{k}}|
\psi_{q\mathbf{k}} \rangle = \sum_{\mathbf{G}}
C_{q^\prime\mathbf{k}}(\mathbf{G})C_{q\mathbf{k}}(\mathbf{G})=\delta_{q^{\prime
}q}. \label{Cqk_bi-expansion}
\end{equation}
It can be seen that 
$\sum_{\mathbf{G}} C_{q\mathbf{k}}^{2}(\mathbf{G}){=}1$, as
opposed
to the usual expression 
$\sum_{\mathbf{G}}
\left|C_{q\mathbf{k}}(\mathbf{G})\right|^2{=}1$ (Parseval's
theorem).
%
%
Note that for the lattices with inversion symmetry discussed here, the two 
sums 
are 
consistent 
in the limit $ \text{Im} (\omega_{q\mathbf{k}}) {\rightarrow} 0
$ (Hermitian Hamiltonian case).
In this limit, $\tilde{\psi}_{q\mathbf{k}}{\rightarrow}{\psi}_{q\mathbf{k}}$ 
and the phases of wave functions
can be adjusted 
to obtain real expansion coefficients 
$C_{q\mathbf{k}}(\mathbf{G})$.

Thus, in 
photonic structures
exhibiting inversion symmetry of the lattice, 
the biorthonormal 
plane wave expansion series 
differ from the usual OPW series 
by normalization condition
for the amplitudes of spatial harmonics. 
The 
Schr\"odinger equation  (\ref{H_all}) can be then readily
converted into a matrix equation 
for $C_{q\mathbf{k}}(\mathbf{G})$ coefficients.
Furthermore, the inversion symmetry of the lattice 
assumes
a simple
relationship between the 
the matrix elements 
of operators 
$\hat{H}$
and
$\hat{H}^{+}$.
%
%
%
Thus, the periodic 
crystal potential
and its adjoint operator
are 
represented by 
series of lattice harmonics 
with complex conjugated coefficients
%
%
\begin{equation}
\begin{split}
{U(\mathbf{r}_\perp)}&{=}
{\Bigl\langle} \hspace{-0.01in} \frac 1n \hspace{-0.01in}
{\Bigr\rangle} 
{\Bigl[}\frac{ic\hbar }{2L} {\ln}
{R}{(}\mathbf{r}_{\perp}{)} 
{+} m_{0}c^2
{\Bigl(}
\frac 1{n{(}\mathbf{r}_\perp{)}} {-} {\Bigl\langle}
\hspace{-0.01in} \frac1n \hspace{-0.01in} {\Bigr\rangle}
{\Bigr)} {\Bigr]}\\
&{=}\sum_{\mathbf{G}} V_\mathbf{G} \exp(-i\mathbf{Gr}_\perp),\\
{U(\mathbf{r}_\perp)^{+}}&{=}U(\mathbf{r}_\perp)^{*}
{=}{\sum_{\mathbf{G}}} V_\mathbf{G}^{*} \exp(-i\mathbf{Gr}_\perp),
\end{split}
\label{U&Ucc_Fourier}
\end{equation}
where 
$V_\mathbf{{-}G}{=}V_\mathbf{G}$ 
due to the symmetry of crystal potential 
$U({-}\mathbf{r}_{\perp}
)=U(\mathbf{r}_{\perp}
)$. 
Substituting 
(\ref{Psi_OPWs}) and 
(\ref{U&Ucc_Fourier}) into the Schr\"odinger equation
(\ref{H_all}) and its $PT$-transform,
multiplying by $\exp(-i(\mathbf k_\perp +
\mathbf G^\prime) \mathbf r_\bot )$
and integrating over 
a lattice cell in the $x$-$y$ plane, one obtains the matrix 
equations
\begin{equation}
\begin{split}
\sum_{\mathbf{G}} \Bigl[ \Bigl(m_{0}c^2
 \Bigl \langle  \frac 1n
 \Bigr \rangle^{2}
& {+} \frac{\hbar^2(\mathbf{k_{\bot }{+}G})^{2}}{2m_{0}}
  {-}\hbar
\omega _{q\mathbf{k}} \Bigr)
\delta_\mathbf{G^\prime G} \Bigr. \\
&  {+}  \Bigl. V_\mathbf{G-G^\prime}\Bigr ]
C_{q\mathbf{k}}(\mathbf{G}){=} 0
\\ 
\sum_{\mathbf{G}} \Bigl[ \Bigl(m^{*}_{0}c^2
 \Bigl \langle \hspace{-0.03in} \frac 1{n^{*}}
 \hspace{-0.03in} \Bigr \rangle^{2}
& {+} \frac{\hbar^2(\mathbf{k_{\bot }{+}G})^{2}}{2m^{*}_{0}}
  {-}\hbar \omega^{*} _{q\mathbf{k}} \Bigr)
\delta_\mathbf{G^\prime G} \Bigr. \\
&  {+}  \Bigl. V^{*}_\mathbf{G-G^\prime}\Bigr ]
C^{*}_{q\mathbf{k}}(\mathbf{G}){=} 0.
\end{split}
\label{H_GG}
\end{equation}
%
%
These
are the 
two 
complex conjugated  matrix eigenproblem equations. 
%
Within the accuracy of complex eigen values and normalization
condition for expansion coefficients, the eigenproblem
(\ref{H_all}) is thus converted into the usual form encountering
in conventional OPW expansion method.

\subsection{Band structure of square-lattice arrays of optical microcavities or parallel waveguides}
\label{NumRes}

To study 
the light propagation behaviour 
in 
paraxial photonic structures incorporating loss and gain
distributions,
the non-Hermitian Hamiltonian
(\ref{H_all}) 
is used 
here to analyze 
the structures with simple lattice symmetry
and
 cell topology.
%
%
Its numerical solutions
are reported 
for 
square-symmetry 
lattices 
depicted 
in Figs.\ref{fig1}
and \ref{fig2}. 
The results 
apply both to the structures  
defined 
by mirror reflectivity  patterning
(\emph{e.g.}, 
metal-patterned VCSEL arrays\cite{Orenstein91}) 
and
to the ones exhibiting 
periodic variations of refractive index (\emph{e.g.}, etched
VCSEL arrays\cite{Gourley91} or matrices of antiguided VCSEL
microcavities\cite{Mawst03}). 

%

All matrices of microcavities 
considered 
here have similar
lattice cell topology 
indicated in Figs.\ref{fig1} and
\ref{fig2}.
%
%
%
%
%
%
%
%
In the case of cavities with mirror reflectivity patterning,
the position of optical microresonators
is 
defined by high-reflectivity 
square pixels
separated by low-reflecting cavity mirror domains forming a grid
pattern.
%
%
The 
fill factor of such square lattice 
is 
the area ratio $FF{=}a^2/\Lambda^2$, with $a$ being
the square pixel width and $\Lambda$ being the lattice
pitch.
%
%
%
In 
similar dielectric lattices defined by 
refractive index variations, 
the position of microcavities is set 
by 
square 
dielectric waveguide (antiguide) cores embedded in the background
of the
cladding material. 
Obviously, the same 
expression
for the 
fill factor 
can be used to characterize 
such 
lattices of dielectric waveguides, 
with $a$ being the %
%
waveguide (antiguide) core width.



%

The Hamiltonian (\ref{H_all}) and the gauge transformation
(\ref{Gauge_all}) utilize 
equivalent, cavity-unfolded (3D) representation of 
microcavities. 
%
%
%
%
%
As discussed in Sec. \ref{GenHamiltonian},
the longitudinal
component 
of propagation vector $\mathbf{k}$ is defined 
by condition $k_z{=}2\pi{/}L$ (in case of one-wavelength
microcavities).
%
%
%
%
Furthermore, by virtue of the dissipative effects rendering
the opposite $z$-axis directions nonequivalent,
%
%
%
the group of 
$\mathbf{k}$ in Eq.~(\ref{Gauge_all}) contains only symmetry
operations preserving
the 
$z$-axis direction.
%
%
%
%

In the particular case of square lattice, 
the
cavity-unfolded 3D
structure 
($2L$-periodic in the $z$-axis direction)
has 
a tetragonal 
symmetry 
($\Lambda{\neq}2L$)
%
with 
symmorphic space group $\Gamma_q$
associated with $D_{4h}$ point subgroup of
rotations.\cite{LandauIII,LandauV, InuiTanabe}
%
%
%
The reciprocal lattice is of the tetragonal symmetry as well with
the first Brillouin zone (BZ) being of the rectangular 
prism shape
[Fig.\ref{fig1}(e)].
%
%
%
%
However, 
the crystal
 $z$-axis nonreciprocity 
implies that
rotations and reflections of the $D_{4h}$ group altering the
$z$-axis direction
are not allowed. 
%
%
%
Therefore, for square arrays of microcavities, 
the 
group 
of $\mathbf{k}$ 
has 
a reduced symmetry characterized by 
$C_{4v}$ point
group rotations in the $\Delta$  and $T$
points of the BZ [at $\mathbf{k}{=}(0,0,k_z)$ and
$(\pm\frac{\pi}{\Lambda},\pm\frac{\pi}{\Lambda},k_z)$,
respectively] and
 $C_{2v}$ symmetry in the $Z$ point
[at $\mathbf{k}=(\pm\frac{\pi}{\Lambda},0,k_z)$ or
$(0,\pm\frac{\pi}{\Lambda},k_z)$].
%
%
%
%
%
%
%
%
%
%
%
%
%
%
%
%

\begin{figure}[tbp]
\includegraphics{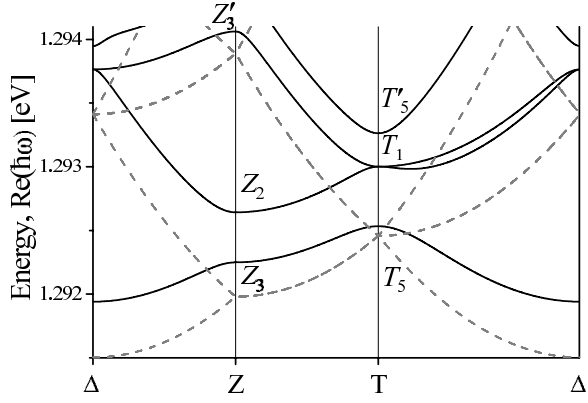}
\caption{ Photon energy in  square-lattice array of microcavities
defined by mirror reflectivity patterning with the phase contrast
of $\delta \ln R= i 10^{-2}$ (solid curves). Dashed curves
indicate band structure of empty lattice ($\ln R= 0$). Other
parameters are $\Lambda=5~\mu m$, $FF=0.5$, $n=3.5$, $L=266 nm$
(one-wavelength cavity optimized for $\lambda=960~nm$). }
\label{fig3}
\end{figure}

\begin{figure}[tbp]
\includegraphics{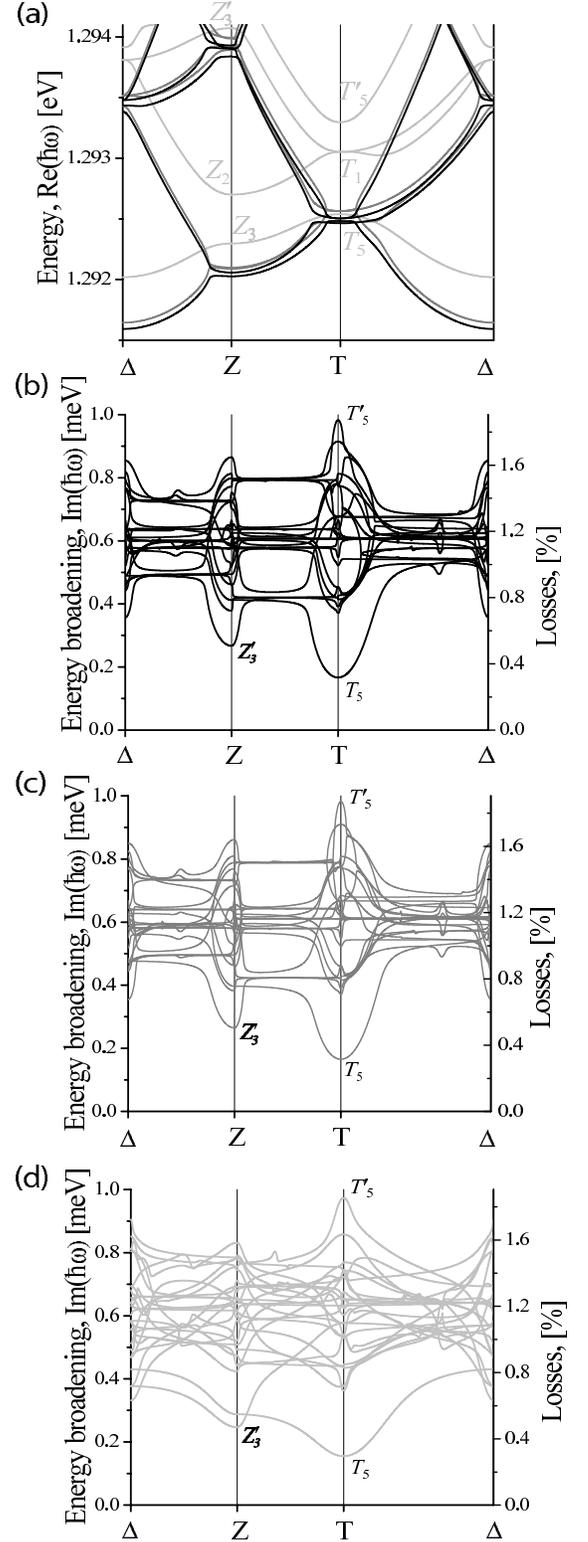}
\caption{Photon energy (a) and lifetime energy broadening (b)-(d)
in array of microcavities with reflectivity patterning
contrast $\delta \ln  R$ of $10^{-2}$ (black curves)
$10^{-2}+i10^{-3}$ (gray curves) and $10^{-2}+i10^{-2}$ (light
gray curves). Other parameters of the structures are listed in the
caption 
of Fig.\ref{fig3}.
}
\label{fig4}
\end{figure}

\begin{figure}[tbp]
\includegraphics{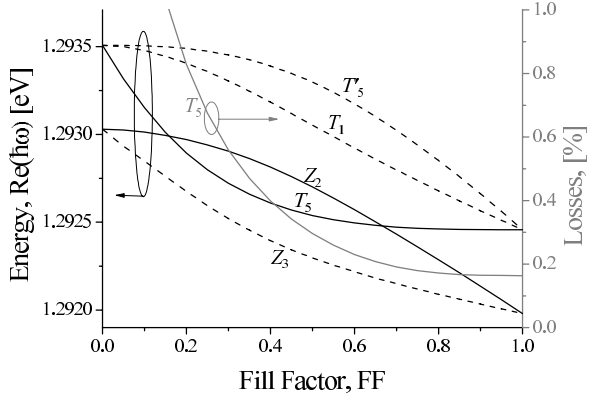}
\caption{Double photonic crystal band gap: Band edge energies of the $T_5$ and $Z_2$ states 
(left axis, black curves) and optical loss of the $T_5$ state
(right axis, gray curve) as a function of the pattern fill factor
FF
in array of microcavities with reflectivity patterning
contrast $\delta \ln R= (1+i) \times 10^{-2}$. 
Other parameters of the structure are listed in the
caption 
of Fig.~\ref{fig3}. The double photonic crystal band gap is
opened in the range of $0.16{<}FF{<}0.66$.
} \label{fig5}
\end{figure}

Figs.\ref{fig3}-\ref{fig5} show 
the results of 
band structure
computations 
for 
square 
lattices of microcavities with the pitch $\Lambda=5~\mu m$ and
lattice cell fill factor $FF=0.5$.
The photon
energy dispersion and lifetime energy broadening 
curves along the 
high-symmetry lines
$\Delta$-$Z$-$T$-$\Delta$ in the BZ
 are obtained here 
using
25 lattice harmonics in Eq.(\ref{H_GG}).
%
%
%
In numerical simulations, the cavity structure is 
assumed to be optimized for $960~nm$ wavelength (
operation wavelength range of GaAs/AlGaAs VCSEL structures with
InGaAs quantum wells in the optical gain region).
%
%

Figs.~\ref{fig3} and \ref{fig4} detail 
the impact of mirror
patterning
contrast
%
%
$\delta \ln R= \delta |R| +i \delta \varphi $
and take contributions of the periodic 
phase ($\delta \varphi$) and amplitude ($\delta |R|$) variations
of the mirror
reflectivity 
into account.

Fig.~\ref{fig3} shows the 
energy dispersion curves for 
low-oder 
photonic bands 
in the case
of a lattice 
defined by
phase modulation of mirror reflectivity with 
parameters $\delta |R|{=}0$ and $\delta \varphi{=}10^{-2}$ (black
curves).
%
%
%
%
%
%
The 
reflectivity at pixel positions is $R=1$ 
while the 
periodic 
pattern 
$R(\mathbf{r}_\perp)$ 
is
introduced by a phase shift $-\delta \varphi$ at reflections form
the cavity
mirror domains forming a grid. 
%
%
%
Comparison with the band structure of empty lattice
(Fig.\ref{fig3}, gray curves) indicates 
%
a blueshift of the optical modes due to contribution of phase-advancing domains of the 
grid. 
The main out-of-phase mode $T_5$ (doubly degenerate by polarization)\cite{Guerrero04B} has 
maxima of probability amplitude $|\psi|$ located at the pixel
positions 
and zeros 
located at the grid. Therefore, it
shows the smallest energy shift. 
The contribution of the grid 
is particularly
pronounced for $T^{\prime}_5$ states exhibiting 
$|\psi|$ distribution with maxima located at the cross points of
the
grid and zeros at the pixels.

Numerically calculated
wave functions of these 
states (not shown in the
figures) 
are in good quantitative 
agreement with 
the results\cite{Boiko07} obtained by means of group theoretical
analysis.\cite{Boiko06D}
%
%
%
The intensity patterns of the optical modes associated with these
states are considered in Refs.~\onlinecite{Boiko02,Boiko04} both
theoretically and experimentally.

%
%
%
%
%
%
%
%
%
%
%
%
%
%
%

Even for such 
low-contrast 
reflectivity patterns as 
the ones considered here, 
a complete 2D band gap can be opened in the lateral direction, 
between the $T_5$ and $Z_2$ states. Thus, in Fig.\ref{fig3}, 
the gap in the 
spectrum of 
optical modes 
is of $0.1~meV$ width.
Since the effect of the 
grid on 
$T_5$ states is small, 
the width of the 
energy gap 
is mostly defined by the blueshift 
of $Z_2$ states 
of an empty lattice.

Wave functions of the $Z_2$ and $Z_3$ states 
in the first BZ have
 zeros 
only along one crystalline 
direction (either $x$ or $y$-axis directions) and differ by the
position of probability amplitude $|\psi|$ maxima. Along this 
lattice direction, the $Z_3$ states are localized to the pixels
while $Z_2$ states are localized to the grid, like, respectively,
$T_5$ and $T^{\prime}_5$ states. However, wave functions of the
states originating from the $Z$ points in the first BZ show large
probability amplitude oscillations (and localization) along one
lattice direction, as opposed to the states from the $T$ point
with wave functions oscillating
in both lateral crystalline directions.
%
%
%
For completeness of the discussion note that the wave functions of
the lowest energy
states 
in the $\Delta$ point (doubly degenerate by polarization
$\Delta_1$ states) show 
no zeros
of probability amplitude $|\psi|$. 
%
Therefore, due to the effect of the grid
pattern, the $Z_3$ ($Z_2$) states 
exhibit 
an intermediate blueshift 
as compared with the 
energy shifts of 
the
$\Delta_1$ 
and  $T_5$ states ($\Delta_1$ and $T^{\prime}_5$ states,
respectively)  of an empty lattice.
%
%
%


Fig.~\ref{fig3} shows 
other states in the BZ ($Z^{\prime}_3$ states) 
exhibiting 
small energy blueshift, 
which is
comparable to that one of the 
$T_5$ states with wave functions localized to the pixel positions.
The $Z^{\prime}_3$ states originate 
from the next nearest
equivalent
$Z$-points of reciprocal lattice 
at $(\pm \frac{\pi}{\Lambda},\pm 2\frac{\pi}{\Lambda},k_z)$ and
$(\pm 2\frac{\pi}{\Lambda},\pm \frac{\pi}{\Lambda},k_z)$.
Therefore, in addition to the unidirectional oscillation features
of
the 
$Z$-states in 
the first BZ, their wave functions 
show large oscillations of probability amplitudes in the second
crystalline direction as well. 
The probability densities of the $Z^{\prime}_3$  states thus have
a better overlap 
with the pixels of reflectivity pattern, which explains 
the
smaller blueshift energy of $Z^{\prime}_3$ states as compared to
their counterparts in the first BZ. In Fig.~\ref{fig3}, the energy
broadening of the bands due to the optical cavity loss has not
been taken into account.


Fig.\ref{fig4} illustrates the effects of 
optical loss distributions in the structures with the same lattice
parameters as in Fig.{\ref{fig3}} ($\Lambda{}=5~\mu m$ and
$FF{=}0.5$).
All structures in Fig.~\ref{fig4} have 
the same amplitude reflectivity patterning $|R(\mathbf{r}_\perp)|$
and differ only by the phase contrast of the 
pattern. The amplitude reflectivity $|R(\mathbf{r}_\perp)|$ is
$0.999$ and $0.989$ for the pixels and grid domains of the
pattern, respectively, yielding the amplitude contrast of $\delta
|R|{=}10^{-2}$. Note that the pixel reflectivity corresponds to
the cavity loss of
$0.2\%$, which is in the range of optical losses in a typical
VCSEL structure. As in the case of the structures in
Fig.~\ref{fig3}, in Fig.~\ref{fig4}, reflections at pixel domains
of reflectivity pattern introduce no additional phase shift into
the cavity roundtrip phase accrual of optical modes. Variations
in the photon energy and lifetime broadening dispersion curves in
Fig.~\ref{fig4} are thus introduced by different
phase-advancing shifts at reflections from the grid domains
in these structures.


At no 
phase modulation of 
reflectivity pattern
[$\delta \varphi=0$],
the 
energy band structure 
is close to the one of an empty lattice but
also shows a set of new peculiar features
[Figs.\ref{fig4} (a), 
black curves]. Thus the degeneracy in energy of photonic states in 
high symmetry points of the BZ is partially removed (the states
are doubly degenerate by polarization) and a set of partial flat
bands appears in the $Z$-
and $T$-points of the BZ. 
These features of 
energy dispersion curves are uniquely defined 
by the dissipative effects 
in the structure.
%
%

The 
energy broadening curves (or loss-dispersion curves) 
of 
optical modes 
in this structure are shown
in Fig.~\ref{fig4} (b). (Only the losses related with the
disperssive features of the patterned cavity mirror are taken into
account.) The doubly degenerate photonic states $T_5$
have the lowest cavity loss (and lifetime energy broadening).
They are of particular 
interest
since they define 
the main lasing modes in coupled laser
arrays.\cite{Boiko04,Guerrero04B} The $Z^{\prime}_3$ states
define the next lowest-loss modes 
in the high-symmetry points of reciprocal lattice. Finally, the
states $T^{\prime}_5$ are associated with the highest-loss optical
modes. 
%
%
%
Using the same considerations as in the case of 
energy
dispersion curves 
in Fig. \ref{fig3}, one can readily
explain these features of the 
loss dispersion curves 
in terms of the overlap between 
photonic state wave functions and high reflectivity pixels.


Introduction of 
phase variations at the array 
grid 
does not affect
the optical losses of 
modes associated with 
$T_5$, $Z^{\prime}_3$ and $T^{\prime}_5$
states [see Figs.\ref{fig4} (b), (c) and (d)].
Fig.\ref{fig4} shows the energy- and loss-dispersion curves 
calculated for the 
phase variations contrast
$\delta \varphi$ of $10^{-1}$ (gray curves)
and $10^{-2}$ (light gray curves). Note that in the last case
$\delta \varphi{=} \delta |R| $.
For 
small phase variations 
$\delta \varphi{<} \delta |R|$ (gray curves),
the structure of energy dispersion bands is close
to that one in the case of pure amplitude modulation of
reflectivity
pattern 
(see Fig.\ref{fig4} (a), gray and black curves). However, for
$\delta \varphi{\sim} \delta |R| $ (light
gray curves), it approaches the energy band structure 
in the case of pure phase modulation of mirror reflectivity
(Fig.\ref{fig3}, black curves) and exhibits a band gap between
$T_5$ and $Z_2$ 
states.

%
%
%
%
%
%

A common feature of the loss-dispersion curves in Fig.~\ref{fig4}
is the opened photonic band gap in the loss
domain\cite{Guerrero04} (or in the domain of photon lifetime in
the cavity). Thus, it is
impossible to excite an optical mode showing 
longer cavity lifetime than in the $T_5$ state. By properly
varying 
the lattice cell fill factor,\cite{Guerrero04,Lundeberg05} this
effect has been used to implement photonic crystal
heterostructure wells capable to confine photonic envelope wave
functions to the regions of lower band gap material.
The confined states show usual features with cosine envelope
functions in the well and exponential tails in the cladding
material. The dissipative photonic crystal materials
have been used so far in 
such photonic crystal
heterostructures
show 
no band gaps in the photon energy domain.

For a structure with complex parameter of reflectivity patterning
contrast $\delta \varphi {=} \delta |R|{=}10^{-2}$
(the same parameters as in Fig.~\ref{fig4}, light gray curves),
Fig.~\ref{fig5} shows variations of the two
photonic band gaps (in the photon energy domain 
and
in the optical loss spectrum) 
as a function of the lattice cell fill factor $FF$. The lowest
loss state $T_5$
(
gray curve, right axis) defines the band edge in the cavity loss
domain. The gap below this edge is opened at any fill factor of
the lattice. 
In the range of lattice cell fill factor $0.16-0.66$, the energy
of $Z_2$ state is higher than the band edge $T_5$, such that a
second band gap originating at the band edge $T_5$ exists in the
photon energy domain, in parallel 
with the gap in the
photon lifetime domain.

The notion of double photonic crystal band gap 
illustrated 
in Fig.~\ref{fig5}
%
opens new 
possibilities for tailoring 
photonic envelope wave function and controlling quantization
features of confined photonic states in photonic crystal heterostructures. 
Thus, 
the  eigenvalues of the Hamiltonian
(\ref{H_all}) 
at a 
photonic crystal heterostructure barrier
assume the dispersion relationship
\begin{equation}
\hbar \omega_A + \frac{\hbar^2 \mathbf{k}^2_{\perp,A}}{2m_A}=\hbar
\omega_B + \frac{\hbar^2 \mathbf{k}^2_{\perp,B}}{2m_B},
\label{Complex_hetero}
\end{equation}
where indexes $A$ and $B$ distinguish photonic crystal materials
at the heterostructure barrier and the parabolic band
approximation
is used for both photonic crystal materials. 
For complex band edge parameters 
$\hbar \omega_{A,B}$ 
and effective masses $m_{A,B}$, this condition assumes that both
propagation constants $\mathbf{k}_{\perp,A}$ and
$\mathbf{k}_{\perp,B}$ are complex, independent of particular 
photonic crystal heterostructure configuration.

Thus, for an $N$-dimensional photonic crystal heterostructure
well, Eq.(\ref{Complex_hetero})
envisages a possibility of $2N$-dimensional confinement of
photonic envelope
wave functions by introducing 
quantization of both real and imaginary parts of propagation
vector $\mathbf k_\perp=\mathbf k^{\prime}_\perp+i\mathbf
k^{\prime\prime}_\perp$ of confined photonic states.
For such states, $\mathbf{k}^{\prime\prime}_\perp \neq 0$ even in
the region of lower band gap material (at the well core), 
allowing the confined 
states to be excited 
at the energies \emph{within forbidden energy gaps} of the
heterostructure 
materials (well core and barrier materials). 

In the numerical examples of band structure computations
presented
here,
photonic structures 
utilizing reflectivity patterning for definition of periodic
crystal lattice are considered. 
The effective crystal potential of these structures is governed
by the third term in the Hamiltonian of Eq.~(\ref{H_all}) and the
magnitudes of its matrix elements are bounded 
to the second-oder perturbations (of the relative order $\sim
\xi^2$).
Therefore, in considered here case of 
paraxial light propagation with photon energy of 
$1.3~eV$ and 
$\xi^2{\sim}10^{3}$, the Hamiltonian (\ref{H_all}) yields 
accurate estimates of
photonic bands 
splitting up to $ 1 ~ meV $.
%
%
%
%
At the same time, 
for dielectric lattices defined by periodic variations of
refractive index, the effective crystal potential [fourth term in
the Hamiltonian]
is bounded to the first-order perturbations ($\sim\xi$). In the
case considered here
($\hbar \omega = 1.3 ~eV$, $\xi{\sim}0.03$),
the Hamiltonian (\ref{H_all}) allows 
energy dispersion curves with band structure 
splitting up to $ 40 ~ meV $ to be analyzed. 

\section{ Conclusion }

In this paper, a simple
non-Hermitian Hamiltonian formalism is 
developed for a 
subclass of two dimensional photonic crystal structures
characterized by paraxial light propagation in the direction
normal to the 
lattice plane, like in the case of coupled microcavity arrays or
microstructured photonic crystal fibers. It allows the optical
loss or gain distributions to be taken into account in the band
structure analysis 
and envisages the effect of
double photonic band gap
opened both in the photon energy and lifetime domains. 
Predicted novel features of optical mode behaviour 
at double photonic crystal heterostructure barriers with band
edge discontinuities
in the energy and lifetime domains
offer new possibilities for photonic crystal applications in
optoelectronic devices and integrated photonic circuits.

%
%

%
%
%
%





\bigskip

\begin{thebibliography}{99}

\bibitem{Kogelnik71}H. Kogelnik, C.V. Shank,
Appl. Phys. Lett. \textbf{18}, 152 
(1971).

\bibitem{Bykov72} V.P. Bykov,
Zh. Expr. Teor. Fiz. \textbf{62}, 505 
(1972).

\bibitem{Yablonovitch87} E. Yablonovitch,
Phys. Rev. Lett. \textbf{58}, 2059 
(1987).

\bibitem{Akahane03} Y. Akahane, T. Asano, B.-S. Song, S. Noda,
Nature \textbf{425}, 944 
(2003).





\bibitem{Mekis96} A. Mekis, et al.,
Phys. Rev. Lett. \textbf{77}, 3787 
(1996).

\bibitem{John87}S. John,
Phys. Rev. Lett. \textbf{58}, 2486 
(1987).

\bibitem{Foresi97} J.S. Foresi, P.R. Villeneuve, J. Ferrera et all,
Nature \textbf{390}, 143 
(1997).


\bibitem{Yano01} S. Yano, et al.
Phys. Rev. B \textbf{63}, 153316 (2001).

\bibitem{Guerrero04} G. Guerrero, D.L. Boiko, E. Kapon,
Optics Express \textbf{12}, 4922 
(2004).

\bibitem{Russell03} R. Russell, Science \textbf{299}, 358 (2003).

\bibitem{Orenstein91} M. Orenstein, E. Kapon, N.G. Stoffel et al.,
Appl. Phys. Lett. \textbf{58}, 
804 
(1991).



\bibitem{Morgan92} R.A. Morgan, K. Kojima, T. Mullally et al,
Appl. Phys. Lett. \textbf{61}, 
1160 
(1992).

\bibitem{Pier97} H. Pier and E. Kapon,
Opt. Lett. \textbf{22},
546 
(1997).

\bibitem{Pier00} H. Pier, E. Kapon and M. Moser,
Nature \textbf{407}, 880 
(2000).



\bibitem{Lundeberg05} L. D. A. Lundeberg, D.L. Boiko, E. Kapon ,
Appl. Phys. Lett. \textbf{87}, 
241120 (2005).


\bibitem{McGurn93} A. R. McGurn, A.A. Maradudin,
Phys.Rev. B \textbf{48}, 17576
(1993).

\bibitem{Kuzmiak94} V. Kuzmiak, A.A. Maradudin, F. Pincemin,
Phys.Rev. B \textbf{50}, 16835
(1994).

\bibitem{Sigalas95} M.M. Sigalas, C.T. Chan, K.M. Ho, C.M. Soukoulis,
Phys.Rev. B \textbf{52}, 11744 
(1995).

\bibitem{Boiko06C} D. L. Boiko, P. F{\'e}ron, and P. Besnard, 
Phys. Rev. B \textbf{73}, 035204 (2006).


\bibitem{Boiko07} D.L. Boiko, "Coriolis-Zeeman effect in rotating photonic
crystal," arXiv:0705.1509 (http://arxiv.org/abs/0705.1509), May
(2007)


\bibitem{Boiko02} D. L. Boiko, G. Guerrero, and E. Kapon,``Bloch wave states
in photonic crystals based on VCSEL arrays",
Proceedings of the 26th International Conference on the Physics
of Semiconductors, ICPS 2002, Edinburgh, 29 July -- 2 August
2002, Institute of Physics Conference Series Number 171,
Institute of Physics Publishing, Bristol (UK), P278 (2003).
http://www.icps2002.org

\bibitem{Boiko04} D.L. Boiko, G. Guerrero, E. Kapon,
Optics Express \textbf{12}, 2597 
(2004).


\bibitem{Guerrero04B} G. Guerrero, D.L. Boiko, and E. Kapon,
Appl. Phys. Lett. \textbf{84}, 3777 
(2004).

\bibitem{Boiko08} D.L. Boiko, (manuscript in preparation).

\bibitem{Lundeberg07}L. D. A. Lundeberg, D. L. Boiko, E. Kapon,
IEEE J. Sel. Top. Quantum Electron. \textbf{13}, 1309 (2007).

\bibitem{MontidiSopra00} F.Monti di Sopra, M. Brunner, H.-P. Gauggel, H.P.
Zappe, M. Moser, R. H\"{o}vel and E. Kapon ,
Appl. Phys. Lett. \textbf{77}, 2283 
(2000).

\bibitem{Boiko06B} D. L. Boiko, G. Guerrero, and E. Kapon,
J. Appl. Phys. \textbf{100}, 103102 (2006).

\bibitem{Leung90} K.M. Leung, Y.F. Liu,
Phys. Rev.B \textbf{41} , 10188 (1990).


\bibitem{LandauI} L.D. Landau, E.M. Lifshitz, \textit{Mechanics} (Nauka, Moscow,
1974).


\bibitem{Altmann65} S.L. Altmann, A.P. Cracknell,
Rev. 
Mod. 
Phys. \textbf{37}, 
19 
(1965), S.L. Altmann, C.J. Bradley,
Rev. 
Mod. 
Phys. \textbf{37}, 
33 
(1965).

%
%

\bibitem{Heer64} C.V. Heer,
Phys. Rev. \textbf{134}, A799 
(1964); Proc. of the Third International Conference on Quantum
Electronics 
(Columbia University Press, New York) 
1305 (1963).

\bibitem{Post67}E.J. Post, Rev. Mod. Phys. \textbf{39}, 475 (1967).


\bibitem{Khromykh66} A. M. Khromykh,
Zh. Eksp. Teor. Fiz 
\textbf{50}, 281 (1966).

\bibitem{LandauII} L.D.Landau, E.M.Livshits, \textit{The Classical Theory of
Fields}, (Nauka, Moscow, 1988).


\bibitem{Bolotovskii74} B.M. Bolotovskii and S.N.Stolyarov ,
Sov.Phys.-Usp. \textbf{17}, 875 
(1975) [Usp.Fiz.Nauk \textbf{114}, 569 
(1974)].

\bibitem{Bolotovskii89} B.M. Bolotovskii and S.N.Stolyarov,
Usp.Fiz.Nauk \textbf{159}, 155 
(1989).

\bibitem{LandauVIII} L.D.Landau, E.M.Livshits, \textit{Electrodynamics of
Continuos Media} (Nauka, Moscow, 1992).

\bibitem{Boiko98} D.L. Boiko,
Optics Express \textbf{2}, 
397 
(1998). 


\bibitem{Bender98} C.M. Bender, S. Boettcher,
Phys. Rev. Lett. \textbf{80}, 5243 (1998).


\bibitem{Mostafazadeh02} A. Mostafazadeh,
J. Math. Phys.\textbf{43}, 
205 
(2002), A. Mostafazadeh,
J. Math. Phys. \textbf{43}, 
2814 
(2002).

\bibitem{Kogelnik75} H. Kogelnik, \textit{Theory of dielectric waveguides} in
\textit{Integrated Optics}, T. Tamir, Ed., (Springer-Verlag,
New-York, 1975), ch.2.

\bibitem{Kurilko68} V. I. Kurilko,
Radiophysics and Quantum Electronics \textbf{11}, 
696 
(1968) [Izvestiya VUZ. Radiofizika 
\textbf{11} 
1221 
(1968)].

\bibitem{LandauIV} V.B. Berestetskii, E.M. Livshits, L.D. Landau,
\textit{Quantum electrodinamics} (Nauka, Moscow, 1989).

\bibitem{Erikson94} W.L. Erikson and S.  Singh,
Phys. Rev. E \textbf{49}, 5778 (1994)

\bibitem{Note1} Note the relationship 
 $e_{3\alpha\beta} e_{3\mu\nu}{=}
\left|
\begin{smallmatrix}
 \delta_{\alpha\mu}-\delta_{\alpha3}\delta_{\mu3} & \delta_{\alpha\nu}-\delta_{\alpha3}\delta_{\nu3} \\
\delta_{\beta\mu} -\delta_{\beta3}\delta_{\mu3} &
\delta_{\beta\nu} -\delta_{\beta3}\delta_{\nu3}
\end{smallmatrix}
\right|$.

\bibitem {Hadley90}G. R. Hadley,
Opt. Lett. \textbf{15},
1215 
(1990).

\bibitem{Faisal81} F. H. M. Faisal and J. V. Moloney,
J. Phys. B: At. Mol. Phys. \textbf{14}, 3603 
(1981).



\bibitem{Morse53} P.H. Morse and H. Feshbach, \textit{Methods of theoretical
physics, Part 1,} (McGraw-Hill, New York , 1953), Chapt.7
pp.791-895.


\bibitem{LandauIII} L.D. Landau, E.M. Livshits, \textit{Quantum Mechanics -
Non-relativistic Theory} (Nauka, Moscow, 1989).

\bibitem{Luttinger} J.M. Luttinger and W. Kohn,
Phys. Rev. \textbf{97}, 869 
(1955).

\bibitem{Note2} See the expression for $B_0^{n^{\prime}n}$
in Eq.(II-11) of Ref.~\onlinecite{Luttinger}.


\bibitem{Slater46} J.C. Slater, 
Rev. Mod. Phys. \textbf{18}, 441 
(1946).

\bibitem{Gourley91}P.L. Gourley, M.E. Warren, G.R. Hadley et al.,
Appl. Phys. Lett. \textbf{58}, 890 (1991).

\bibitem{Mawst03} L.J. Mawst, 
IEEE Circuits and Devices Magazine \textbf{19}, 34 (2003).

\bibitem{Shteeman07} V.R. Shteeman, D.L. Boiko, E.Kapon, A. A. Hardy,
IEEE J. Quantum Electron. \textbf{43}, 215 (2007)

\bibitem{Argyros05} A. Argyros, T. Birks, S. Leon-Saval, C. M. Cordeiro, F. Luan,
and P. S. J. Russell,
Opt. Express \textbf{13}, 309 (2005).

\bibitem{Wolinski06}T.R. Wolinski, K. Szaniawska, S. Ertman, P. Lesiak, A.W.
Domanski,R. Dabrowski, E. Nowinowski-Kruszelnicki, J. Wojcik,
Meas. Sci. Technol. \textbf{17},  985 (2006).

\bibitem{Bayer99}M Bayer, T Gutbrod, A Forchel, T.L. Reinecke, P. Knipp, R
Werner, J.P.  Reithmaier,
Phys. Rev. Let. \textbf{83}, 
5374 
(1999).

\bibitem{Pekar46} S. Pekar,
Zh. Eksp. Teor. Fiz. 
\textbf{16},  933 (1946).

\bibitem{BonchBruevich90} V.L. Bonch-Bruevich, S.G. Kalashnikov,
\textit{The Phyics of Semiconductors} (Nauka, Moscow, 1990), Chap.
4, pp. 146-151.

\bibitem{Kittel} C.Kittel, \textit{Quantum Theory of Solids} (
Wiley, New York, 1963).

\bibitem{Kramers30} H. A. Kramers, Aliad. Wetenschappen Amsterciam \textbf{33}, 959
(1930).

\bibitem{Wigner32} E. P. Wigner, Nachr. Ges. Wiss. Gottingen, Math.-Physik.
Kl., 546 (1932); E. P. Wigner, \textit{Group Theory} (Academic
Press, New York, 1959).

\bibitem{InuiTanabe} T. Inui, Y. Tanabe, Y. Onodera. \textit{Group Theory and its
Applications in Physics}, (Springer-Verlag, Berlin \textit{etc.},
1989).

\bibitem{LandauV} L.D.Landau, E.M.Livshits, \textit{Statistical Physics, Part 1},
(Nauka, Moscow, 1976).








%





















































\end{thebibliography}
\end{document}